\documentclass[aps,twocolumn,superscriptaddress,prx,amsmath,amssymb,10pt,aps,longbibliography]{revtex4-1} 
\usepackage{graphicx}
\usepackage{latexsym}
\usepackage{amsmath}
\usepackage{wasysym}
\usepackage{amsthm}
\usepackage{amsbsy}
\usepackage{amssymb}
\usepackage{epstopdf} 
\usepackage{enumerate}
\usepackage{setspace} 
\usepackage{dcolumn}
\usepackage{bm}
\usepackage{setspace} 
\usepackage{slashed}
\usepackage{color}
\usepackage{xcolor}
\usepackage[colorlinks=true,linkcolor=teal,citecolor=teal,urlcolor=teal]{hyperref}
\usepackage{float}
\usepackage{graphicx}
\usepackage{multirow}
\usepackage{array}
\usepackage{pifont}
\newcolumntype{L}[1]{>{\raggedright\arraybacKQSLash}p{#1}}
\newcolumntype{C}[1]{>{\centering\arraybacKQSLash}p{#1}}
\newcolumntype{R}[1]{>{\raggedleft\arraybacKQSLash}p{#1}}

\usepackage{color}
\usepackage{makecell}

\renewcommand{\thesubsection}{\normalsize Supplementay~Note~\arabic{subsection}}

\begin{document}

\title{Identification of a Kitaev Quantum Spin Liquid \\ by  Magnetic Field Angle Dependence}

\author{Kyusung Hwang}
	\thanks{These authors contributed equally: Kyusung Hwang, Ara Go.}
	\affiliation{School of Physics, Korea Institute for Advanced Study (KIAS), Seoul 02455, Korea}
	
\author{ Ara Go }
	\thanks{These authors contributed equally: Kyusung Hwang, Ara Go.}
	\affiliation{Center for Theoretical Physics of Complex Systems, Institute for Basic Science (IBS), Daejeon 34126, Korea}
	\affiliation{Department of Physics, Chonnam National University, Gwangju 61186, Korea}
	
\author{ Ji Heon Seong }
	\affiliation{Department of Physics, Korea Advanced Institute of Science and Technology (KAIST), Daejeon 34141, Korea}
	
\author{ Takasada Shibauchi }
	\affiliation{Department of Advanced Materials Science, University of Tokyo, Kashiwa, Chiba 277-8561, Japan}
	
\author{ Eun-Gook Moon }
	\email{egmoon@kaist.ac.kr}
	\affiliation{Department of Physics, Korea Advanced Institute of Science and Technology (KAIST), Daejeon 34141, Korea}

\date{\today}
\begin{abstract}
Quantum spin liquids realize massive entanglement and fractional quasiparticles from localized spins, proposed as an avenue for quantum science and technology. 
In particular, topological quantum computations are suggested  in the non-abelian phase of Kitaev quantum spin liquid with Majorana fermions, and detection of Majorana fermions is one of the most outstanding problems in modern condensed matter physics.  Here, we propose a concrete way to identify the non-abelian Kitaev quantum spin liquid by magnetic field angle dependence. Topologically protected critical lines exist on a plane of magnetic field angles, and their shapes are determined by microscopic spin interactions.  
A chirality operator plays a key role in demonstrating microscopic dependences of the critical lines. We also show that the chirality operator can be used to evaluate topological properties of  the non-abelian Kitaev quantum spin liquid without relying on Majorana fermion descriptions.
Experimental criteria for the non-abelian spin liquid state are provided for future experiments.  
\end{abstract}
\maketitle

A quantum spin liquid (QSL) is an exotic state of matter characterized by many-body quantum entanglement~\cite{Zhou, Balents,Moessner2019}.
In contrast to weakly entangled magnetic states, QSLs host emergent fractionalized quasiparticles described by bosonic/fermionic spinons and gauge fields~\cite{Sachdev, WenBook}.
The exactly solvable honeycomb model by Kitaev reveals the exact ground and excited states featured with Majorana fermions and $\mathbb{Z}_2$ gauge fluxes, so-called Kitaev quantum spin liquid (KQSL)~\cite{Kitaev}.
Strong spin-orbit coupled systems with 4$d$ and 5$d$ atoms such as $\alpha$-RuCl$_3$ are proposed to realize KQSL~\cite{Jackeli2009,Chaloupka2010,Chaloupka2013,Rau2014,KimHS,
KimHS2016,Winter2016,Kee2016ARCMP,Trebst2017,Winter2017review}, and related spin models have been studied intensively~\cite{Baskaran2007,Knolle2014,Perkins,Chaloupka2015,Motome2015,Balents2016,NasuNMR,Perkins2016,Motome2017,
Winter2017magnon_breakdown,Kee2018,YBKim2018,HYLee2019,Batista2019,
Brink2016,Vojta2016,Gohlke2017,Gohlke2018,Fu2018,Winter2018,
Balents2018,Hickey2019,Kee2019,Ronquillo2019,Patel2019,Vojta2019,Chaloupka2019,Takagi2019,Motome_review,YB2020,Wang2019VMC,Wang2020VMC}.

Recent advances in experiments have unveiled characteristics of QSLs. For $\alpha$-RuCl$_3$, signatures of Majorana fermion excitations have been observed in various different experiments of neutron scattering, nuclear magnetic resonance, specific heat, magnetic torque, and thermal conductivity~\cite{Plumb2014, Sears2015, Burch, Nagler, Banerjee2017, Ji, Baek, Buchner, LeeM, Kasahara, Klanjsek, Matsuda, Vojta2018, Wulferding2020, Yokoi2020, Yamashita2020sampledependence, Shibauchi2020specificheat}. 
Among them, the half quantization of thermal Hall conductivity $\kappa_{xy} / T = (\pi/12) (k_B^2/\hbar)$ may be interpreted as the hallmark of the presence of Majorana fermions and the non-abelian KQSL~\cite{Matsuda,Yokoi2020}. At higher magnetic fields, a significant reduction of $\kappa_{xy} / T$ also suggests a topological phase transition~\cite{Matsuda, Go2019}. 
Thermal Hall measurements are known to be not only highly sensitive to sample qualities~\cite{Yamashita2020sampledependence} but also very challenging due to the required precision control of heat and magnetic torque from strong magnetic fields.
This strongly motivates an independent way to detect the Majorana fermions and non-abelian KQSL.

In this work, 
we propose that the non-abelian KQSL may be identified by the angle dependent response of the system under applied magnetic fields. 
As a smoking gun signature of the KQSL, quantum critical lines are demonstrated to occur on a plane of magnetic field directions whose existence is protected by topological properties of the KQSL. 
The critical lines vary depending on the microscopic spin Hamiltonian, which we show by investigating a chirality operator via exact diagonalization.
We further propose that the critical lines can be detected by heat capacity measurements and provide experimental criteria for the non-abelian KQSL applicable to the candidate material $\alpha$-RuCl$_3$.
\\

\begin{figure*}[tb]
\centering
\includegraphics[width=\linewidth]{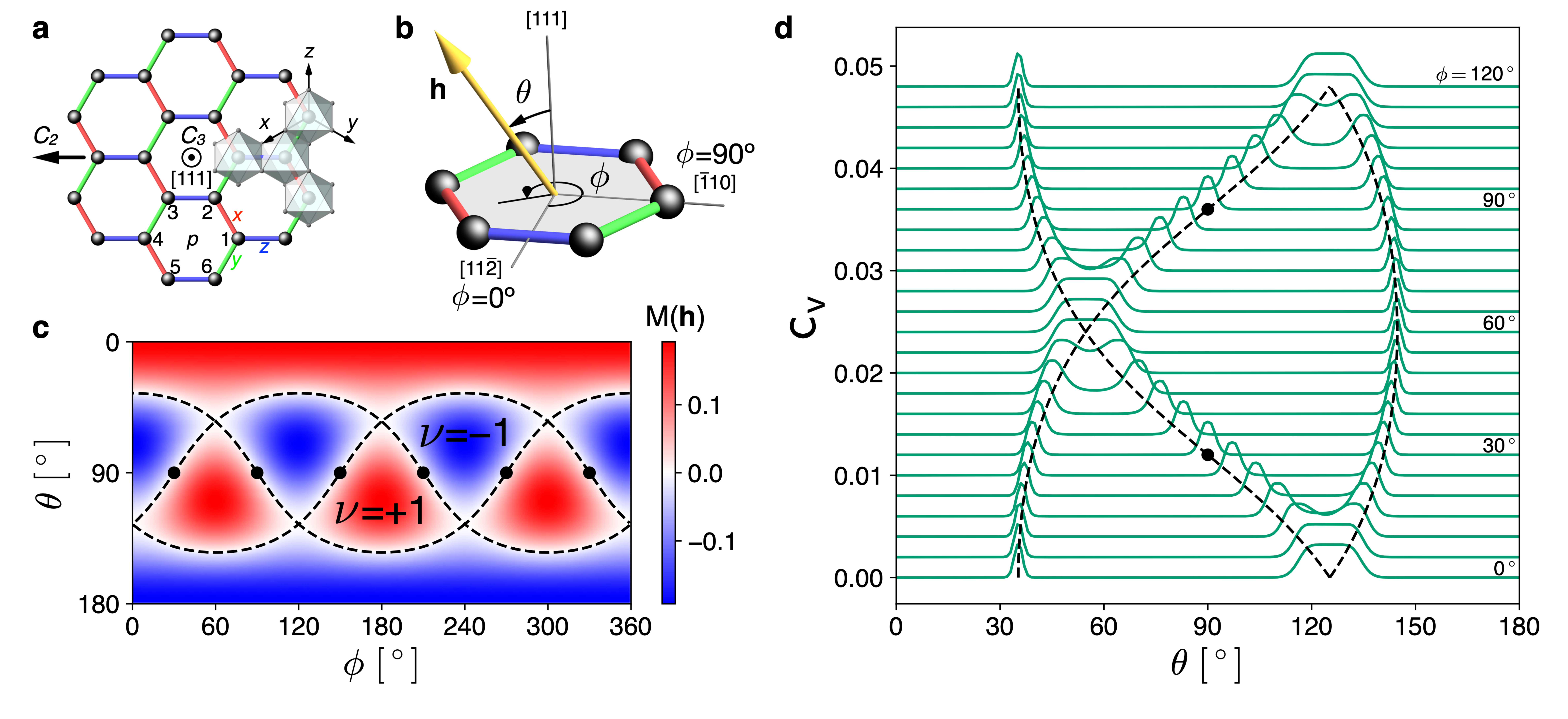} 
 \caption{{\bf Field angle dependence of the pure Kitaev model.} 
{\bf a} Honeycomb lattice enclosed by edge-sharing octahedra. Red, green, blue lines denote the $x$,$y$,$z$-bonds, and 
the six numbers indicate the numbering convention for sites in each hexagon plaquette ($p$).
Black arrows depict $C_3$ and $C_2$ rotation axes.
{\bf b} Convention for the angular representation of an applied magnetic field ${\bf h}$ (yellow). 
The polar and azimuthal angles ($\theta,\phi$) are measured from 
the out-of-plane $[111]$ axis and the bond-perpendicular $[11\bar{2}]$ axis, respectively.
{\bf c} Color map of the mass function $M(\mathbf{h}) = h_x h_y h_z/K^2$ on the ($\theta, \phi$) plane.
The dashed lines highlight the topological phase transitions between the $\nu=\pm 1$ states, i.e., the quantum critical lines of the energy gap $\Delta(\mathbf{h})\propto|M(\mathbf{h})|=0$.
The black dots mark the bond directions 
$(\theta=90^\circ, \phi = 30^\circ+n\cdot60^\circ)$, where $n=0,1,\cdots,5$.
{\bf d} A schematic of the field angle dependence of the heat capacity $c_v$ at a fixed temperature, where peak positions determine the critical lines of the non-abelian KQSL.}
\label{fig:1}
\end{figure*}

\section*{Results}

\noindent{\bf Model and symmetries.}
We consider a generic spin-1/2 model on the honeycomb lattice with edge-sharing octahedron crystal structure,
\begin{eqnarray}
H_{KJ\varGamma\varGamma^\prime} 
=
\sum_{\langle jk \rangle_\gamma} &&
\left[
K S_j^{\gamma} S_k^{\gamma} 
+ 
J \mathbf{S}_j \cdot \mathbf{S}_k 
+ 
\varGamma \big( S_j^{\alpha} S_k^{\beta} + S_j^{\beta}S_k^{\alpha}\big)  
\right.
\nonumber\\
&+& 
\left.
\varGamma^\prime \big( S_j^{\alpha} S_k^{\gamma} +  S_j^{\gamma} S_k^{\alpha} + S_j^{\beta} S_k^{\gamma}+ S_j^{\gamma} S_k^{\beta}  \big) 
\right],
\nonumber
\end{eqnarray}
so-called $K$-$J$-$\varGamma$-$\varGamma'$ model~\cite{Rau2014,KimHS,Winter2016,Winter2017review}.
Nearest neighbor bonds of the model are grouped into $x,y,z$-bonds depending on the bond direction (Fig.~\ref{fig:1}a). 
Spins ($\mathbf{S}_{j,k}$) at each bond are coupled via the Kitaev ($K$), Heisenberg ($J$), and off-diagonal-symmetric ($\varGamma,\varGamma'$) interactions.
The index $\gamma\in\{x,y,z\}$ denotes the type of bond, and the other two $\alpha,\beta$ are the remaining components in $\{ x,y,z\}$ other than $\gamma$.
Under an applied magnetic field (${\bf h}$), the Hamiltonian becomes 
\begin{eqnarray}
H (\theta,\phi)
=
H_{KJ\varGamma\varGamma^\prime}
-
\mathbf{h} (\theta,\phi) \cdot  \sum_{j} \mathbf{S}_j. 
\label{eq:full_H}
\end{eqnarray}
We specify the magnetic field direction with the polar and azimuthal angles $(\theta,\phi)$ as defined in Fig.~\ref{fig:1}b.  
$H_{KJ\varGamma\varGamma'}$ possesses the symmetries of time reversal, spatial inversion, $C_3$ rotation about the normal axis to each hexagon plaquette, and $C_2$ rotation about each bond axis (Fig.~\ref{fig:1}a).
The $C_3$ and $C_2$ rotations form a dihedral group $D_3$.
Under each of these symmetries, $H (\theta,\phi)$ is transformed to $H (\theta',\phi')$ with a rotated magnetic field $ \mathbf{h} (\theta',\phi')$; see Supplementary~Notes~1~and~2.

In the pure Kitaev model, parton approach provides the exact wave function of KQSL together with gapped $\mathbb{Z}_2$ flux and gapless Majorana fermion excitations.
Application of magnetic fields drives the KQSL into the non-abelian phase by opening an energy gap in Majorana fermion excitations. 
The gap size is proportional to the mass function, $M(\mathbf{h}) = h_x h_y h_z/K^2$, and the topological invariant (Chern number) of the KQSL is given by the sign of the mass function, ${\rm sgn}(h_x h_y h_z)$~\cite{Kitaev}.
\\

\noindent{\bf Topologically protected critical lines.}
Topological invariant of non-abelian phases with Majorana fermions can be defined from the quantized thermal Hall conductivity, $\kappa_{xy} / T = \nu (\pi/12) (k_B^2/\hbar)$, where $\nu$ is the topological invariant representing the total number of chiral Majorana edge modes ($T$: temperature)\cite{Kitaev}.
While the topological invariant  in the pure Kitaev model is exactly calculated by the Chern number of Majorana fermions, it is a nontrivial task to analyze the topological invariant for the generic model $H(\theta,\phi)$.  

Our strategy to overcome the difficulty is to exploit symmetry properties of the topological invariant and find characteristic features of the non-abelian KQSL. Concretely, we focus on the landscape of $\nu({\bf h})$ on the plane of the magnetic field angles $(\theta,\phi)$. Our major finding is that critical lines of $\nu({\bf h})$ must arise as an intrinsic topological property of the non-abelian KQSL.

We first consider time reversal symmetry and note the following three facts: 
\begin{itemize}
\item Time reversal operation reverses the topological invariant as $\nu (\mathbf{h}) \rightarrow -\nu (\mathbf{h})$.
\item Time reversal operation also reverses the magnetic field direction: ${\bf h}(\theta, \phi) \rightarrow -{\bf h}(\theta, \phi)={\bf h}(\pi- \theta, \phi+\pi)$. 
\item Topologically distinct regions with $\{+\nu (\mathbf{h}), +{\bf h}\}$ and $\{-\nu (\mathbf{h}), -{\bf h}\}$ exist on the $(\theta, \phi)$ plane.
\end{itemize}
These properties enforce the two regions to meet by hosting critical lines where Majorana fermion excitations become gapless. In other words, topological phase transitions must occur as the field direction changes. We propose that the very existence of critical lines can be used in experiments as an identifier for the KQSL.  

We further utilize the $D_3$ symmetry of the system. 
The topological invariant $\nu$ and thermal Hall conductivity $\kappa_{xy}$ are $A_2$ representations of the $D_3$ group, i.e., even under $C_3$ rotations but odd under $C_2$ rotations,
which reveals the generic form,
\begin{eqnarray}
\nu (\mathbf{h}) = {\rm  sgn}[\Lambda_1 (h_x +h_y+h_z) + \Lambda_3 h_x h_y h_z] ,
\label{eq:chern_number}
\end{eqnarray}
where $\Lambda_{1,3}$ are field-independent coefficients. The $h$-linear term $(h_x +h_y+h_z)$ and $h$-cubic term $(h_x h_y h_z)$ are the leading order $A_2$ representations of magnetic fields.
Conducting third order perturbation theory, we find the coefficients
\begin{eqnarray}
\Lambda_1=-\frac{4\varGamma'}{\Delta_{\rm flux}}+\frac{6J\varGamma'}{\Delta_{\rm flux}^2}-\frac{4\varGamma\varGamma'}{\Delta_{\rm flux}^2}+\frac{5\varGamma'^2}{2\Delta_{\rm flux}^2}
~\&~
 \Lambda_3= \frac{18}{\Delta_{\rm flux}^2},~~~~~
 \label{eq:perturbative_parton}
\end{eqnarray}
where $\Delta_{\rm flux}=0.065|K|$ means the flux gap in the Kitaev limit. 
See Supplementary~Notes~3,~4,~5 and Ref.~\cite{Takikawa2019} for more details of the perturbation theory.

Notice that the $h$-linear term is completely absent in the pure Kitaev model ($\Lambda_1=0$).
Figure~\ref{fig:1}c visualizes the topological invariant, $\nu(\mathbf{h})= {\rm  sgn}(h_x h_y h_z)$.
The dashed lines highlight the critical lines representing the topological phase transitions between the phases with  $\nu (\mathbf{h})=\pm1$, where the energy gap of Majorana fermion excitations is closed: $\Delta(\mathbf{h})=0$.

Exploiting the symmetry analysis, we stress two universal properties of the KQSL with the $D_3$ symmetry.
\begin{enumerate}
\item Symmetric zeroes: for bond direction fields, topological transition/gap closing is guaranteed to occur by the symmetry, i.e.,
$\Delta({\bf h})=0$ for 
$(\theta=90^\circ, \phi = 30^\circ+n\cdot60^\circ)$.
\item Cubic dependence: for in-plane fields, the $h$-cubic term governs low field behaviors of the KQSL, e.g.,
$\nu({\bf h})\sim\textup{sgn}(h_x h_y h_z)$ \& $\Delta({\bf h})\sim |h_x h_y h_z|$ for $\theta=90^\circ$.
\end{enumerate}
The universal properties and critical lines of the KQSL are numerically investigated for the generic Hamiltonian $H (\theta,\phi)$ in the rest of the paper.\\

\noindent{\bf Chirality operator.}
We introduce the chirality operator 
\begin{equation}
\hat{\chi}_p
=S_2^xS_1^zS_6^y+S_5^xS_4^zS_3^y
+C_3~{\rm  rotated~terms}
\label{eq:chirality_op}
\end{equation}
at each hexagon plaquette $p$ and investigate the expectation value of the chirality operator, shortly the chirality,
\begin{equation}
\chi (\mathbf{h}) \equiv \frac{1}{N} \sum_p \langle \Psi_{\rm KQSL} (\mathbf{h}) | \hat{\chi}_p | \Psi_{\rm KQSL} (\mathbf{h}) \rangle,
\label{eq:chirality}
\end{equation} 
and its sign,
\begin{equation}
\bar{\nu} (\mathbf{h}) \equiv {\rm sgn} [\chi (\mathbf{h}) ] ,
\end{equation}
where $|\Psi_{\rm KQSL}({\bf h})\rangle$ is the ground state of the full Hamiltonian $H(\theta,\phi)$ in the KQSL phase ($N$ is the number of unit cells). 
The chirality operator $\hat{\chi}_p$ produces the mass term of Majorana fermions and determines the topological invariant in the pure Kitaev limit~\cite{Kitaev}.
More precisely, how magnetic fields couple to the chirality operator determines the topological invariant and the Majorana energy gap.
The chirality $\chi$ and its sign $\bar{\nu}$ are in the $A_2$ representation  of the $D_3$ group as of the topological invariant ${\nu}$.
We note that the relation between the chirality and the Majorana energy gap, $\chi \sim \Delta$, holds near the symmetric zeroes in a generic KQSL beyond the pure Kitaev model due to the symmetry properties (Supplementary~Note~6).
\\

\begin{figure*}[tb]
\centering
\includegraphics[width=\linewidth]{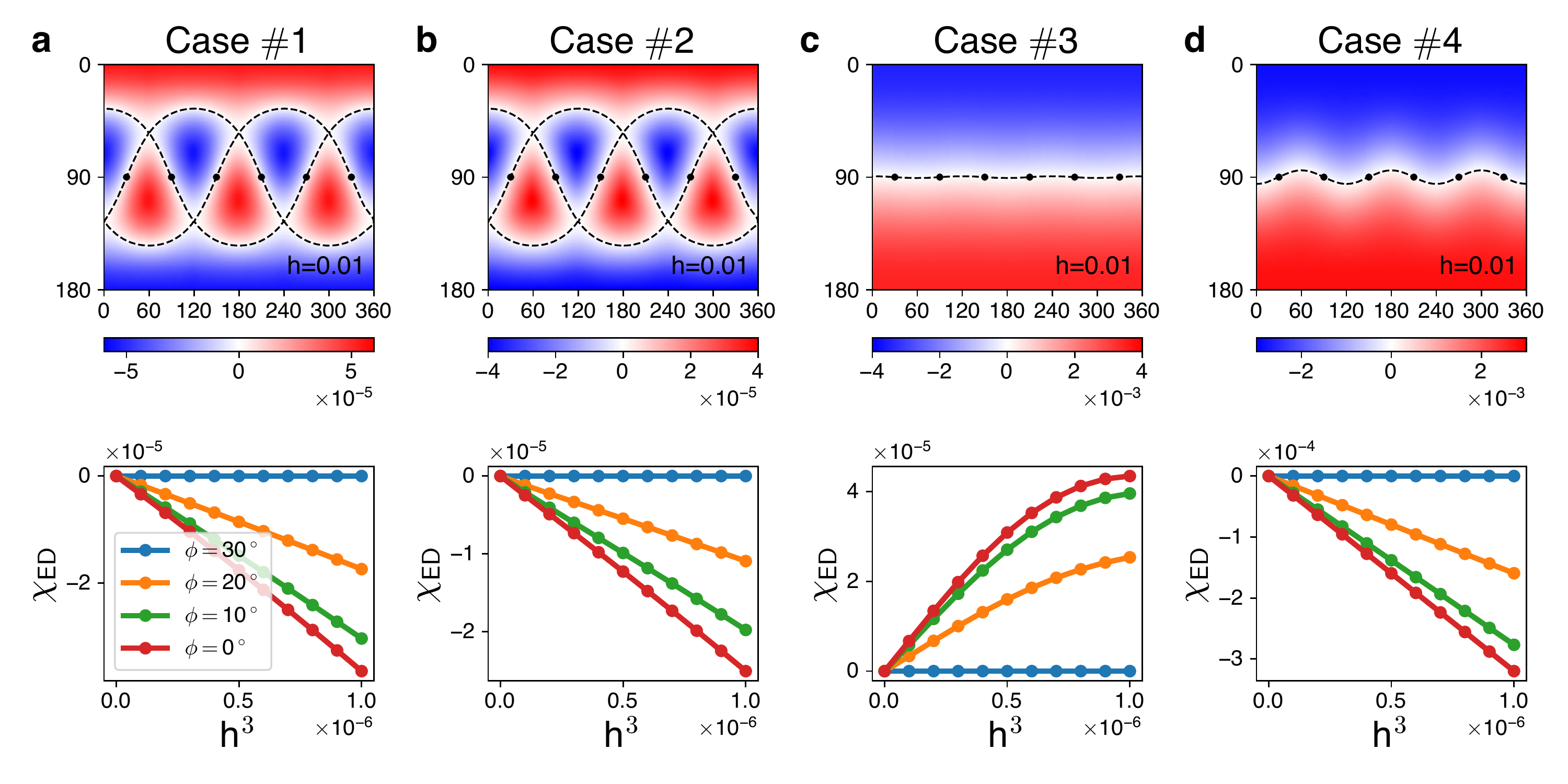} 
\caption{{\bf Chirality of the non-abelian KQSL.}
Upper: color maps of the chirality $\chi_{\rm ED}({\bf h})$ on the plane of the field angles $(\theta,\phi)$, where the magnetic field strength is fixed by $h=0.01$ (horizontal axis: $\phi~[^\circ]$, vertical axis: $\theta~[^\circ]$).
The dashed lines highlight the zero lines $\chi_{\rm ED}({\bf h})=0$, and the black dots mark the bond directions.
Lower: $\chi_{\rm ED}({\bf h})$ as a function of $h^3$ for the in-plane fields $(\theta=90^\circ,\phi=0^\circ,10^\circ,20^\circ,30^\circ)$, illustrating the universality of the $h^3$ behavior in the KQSL.
In the case $\#3$, the bending at $h^3 > 0.5\times10^{-6}$ is an effect of higher order contributions ($h^5,h^7,\cdots$).
The parameter sets used in the four cases ($\#1\sim4$) are listed in Table~\ref{tab:1}.
}
\label{fig:2}
\end{figure*}

\begin{figure*}[tb]
\centering
\includegraphics[width=\linewidth]{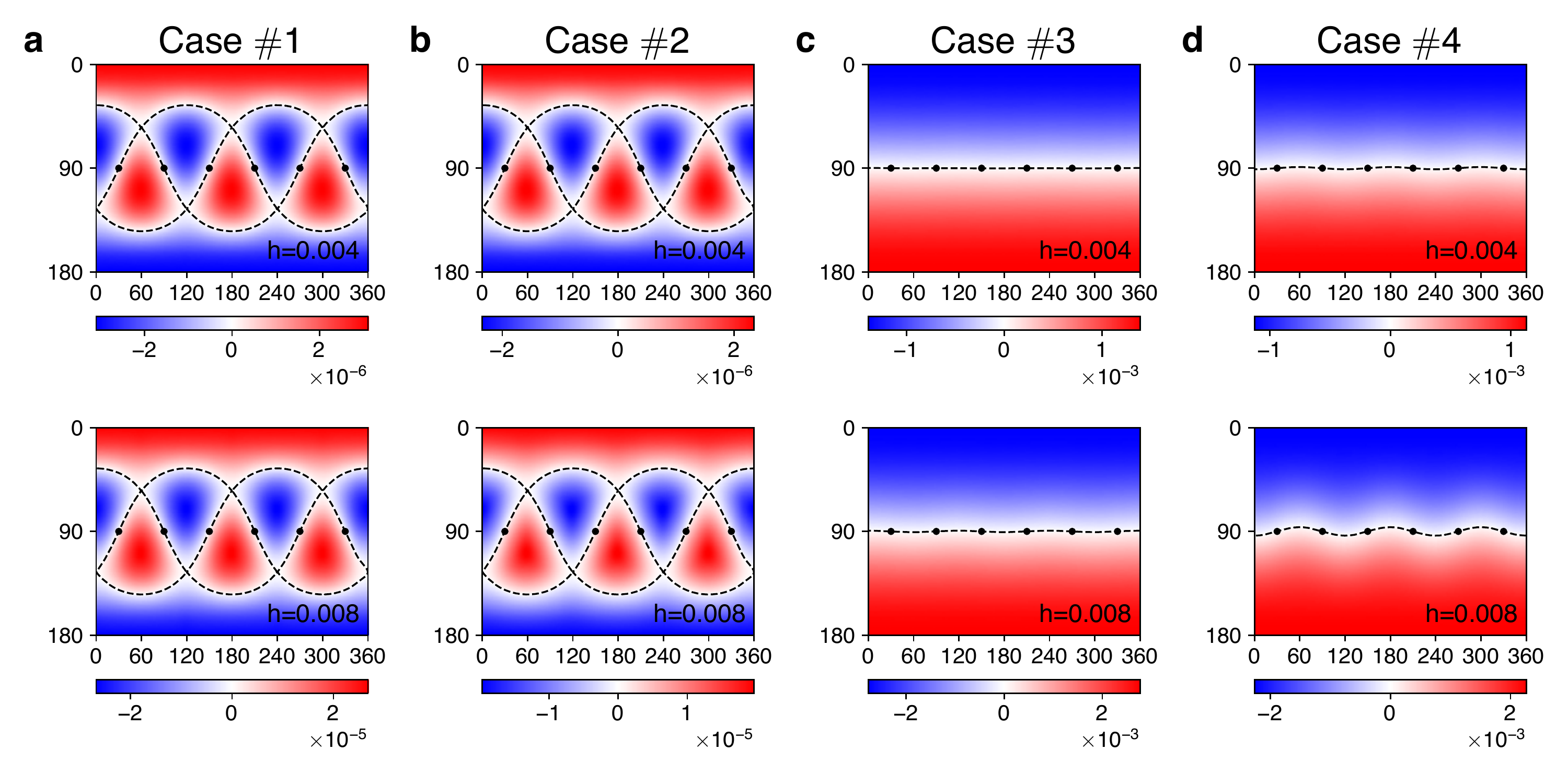} 
\caption{{\bf Field evolutions of the zero lines.}
Color maps of the chirality $\chi_{\rm ED}({\bf h})$ on the plane of the field angles $(\theta,\phi)$ with increasing magnetic field $h=0.004,~0.008$ (horizontal axis: $\phi~[^\circ]$, vertical axis: $\theta~[^\circ]$).
The dashed lines highlight the zero lines $\chi_{\rm ED}({\bf h})=0$, and the black dots mark the bond directions.
The parameter sets used in the four cases ($\#1\sim4$) are listed in Table~\ref{tab:1}.
}
\label{fig:3}
\end{figure*}

\noindent{\bf Exact diagonalization.}
The Hamiltonian $H(\theta,\phi)$ is solved via exact diagonalization (ED) on a 24-site cluster with sixfold rotation symmetry and a periodic boundary condition (Fig.~\ref{fig:1}a).
Resulting phase diagrams are provided in the section of {\bf Methods}.
Figures~\ref{fig:2}~and~\ref{fig:3} display our major results, the ED calculations of the chirality $\chi_{\rm  ED}({\bf h})$ for the KQSL. 
The used parameter sets are listed in Table~\ref{tab:1}.
The zero lines [$\chi_{\rm  ED}({\bf h})=0$; dashed lines in the figures] exist in all the cases.

The two universal features of the KQSL are well captured by the chirality (Figs.~\ref{fig:2}~and~\ref{fig:3}).
Firstly, the zero lines of the chirality $\chi_{\rm  ED}(\mathbf{h})$ always pass through the bond directions (marked by black dots), i.e., the symmetric zeroes.
Secondly, the chirality shows the cubic dependence for in-plane fields ($\theta=90^\circ$).
The linear term, $h_x+h_y+h_z$, vanishes for in-plane fields,
and the cubic term, $h_xh_yh_z$, determines the chirality at low fields, which is confirmed in our ED calculations (lower panels of Fig.~\ref{fig:2}).
Below, we show how  non-Kitaev interactions affect topological properties of the non-abelian KQSL.

\begin{table}[tb]
\begin{ruledtabular}
\begin{tabular}{lcccccc}
Case  &  $K$  &  $J$  &  $\varGamma$  &  $\varGamma'$  &  Phase & Figures
\\
\hline
\#1  & -1  & 0 & 0 & 0 &  KQSL & \ref{fig:2}a,~\ref{fig:3}a
\\
\#2  & -1  & 0.05 & 0 & 0 &  KQSL & \ref{fig:2}b,~\ref{fig:3}b
\\
\#3  & -1  & 0.05 & 0 & 0.05 &  KQSL & \ref{fig:2}c,~\ref{fig:3}c
\\
\#4  & -1  & 0.08 & 0.01 & 0.05 &  KQSL & \ref{fig:2}d,~\ref{fig:3}d
\\
\#5  &  -1  & -0.2 & -0.2 & 0.05 &  FM & \ref{fig:4}b
\\
\#6  &  -1  & 0.2 & 0.05 & 0.05 &  Stripy & \ref{fig:4}c
\\
\#7  &  -1  & 0.2 & -0.2 & 0.05 &  Vortex & \ref{fig:4}d
\\
\#8  &  1  & 0.2 & -0.2 & -0.05 &  Neel & \ref{fig:4}e
\\
\#9  &  -1  & -0.3 & 1 & -0.1 &  Zigzag & \ref{fig:4}f
\end{tabular}
\end{ruledtabular}
\caption{{\bf Parameter sets for exact diagonalization and spin wave calculations.}
}
\label{tab:1}
\end{table}

Most of all, we find that $\bar{\nu}$ becomes identical to $\nu$ for the pure Kitaev model in Fig.~\ref{fig:2}a. 
It is remarkable that the two different methods, ED calculations of the chirality sign and the parton analysis, show the complete agreement: $\bar{\nu}({\bf h})=\nu({\bf h})$. 
The agreement indicates that the topological phase transitions can be identified by using the chirality operator, which becomes a sanity check of our strategy to employ the chirality operator. 

Figure~\ref{fig:2}b illustrates effects of the Heisenberg interaction ($J$) on the chirality. 
The shape of the critical lines is unaffected by the Heisenberg interaction, remaining the same as in the pure Kitaev model.
This result is completely consistent with the perturbative parton analysis [Eq.~(\ref{eq:perturbative_parton}) and Supplementary~Fig.~2b], indicating the validity of our strategy. 

Figure~\ref{fig:2}c-d presents consequences of the other non-Kitaev interactions, $\varGamma'$ and $\varGamma$. The zero lines tend to be flatten around the equator $\theta=90^\circ$, which can be attributed to the $h$-linear term induced by the non-Kitaev interactions: $ h_x +h_y+h_z=   h \sqrt{3} \cos\theta$.
In other words, the zero lines substantially deviate from those of the pure Kitaev model by the non-Kitaev couplings, $\varGamma'$ and $\varGamma$.
We point out  that the signs of the chirality are opposite to the Chern numbers of the third order perturbation parton analysis (Supplementary~Fig.~2c-d). 
The opposite signs indicate that the two methods have their own valid conditions, calling for improved analysis (Supplementary~Note~8).

Impacts of the non-Kitaev interactions also manifest in the field evolution of the zero lines (Fig.~\ref{fig:3}).
Without the non-Kitaev couplings, $\varGamma'$ and $\varGamma$, the shape of the critical lines is governed by the cubic term $h_xh_yh_z$, as shown in Fig.~\ref{fig:3}a-b.
In presence of the $\varGamma'$ or $\varGamma$ coupling, the $h$-cubic term competes with the $h$-linear term as illustrated in Fig.~\ref{fig:3}c-d.
Namely, the linear term dominates over the cubic term at low fields while the dominance gets reversed at high fields (see Supplementary~Note~11 and Supplementary~Table~4 for more results).
The competing nature may be used to quantitatively characterize the non-Kitaev interactions.

Similarities and differences between the topological invariant, $\nu({\bf h})$, and the sign of the chirality, $\bar{\nu}({\bf h})$, are emphasized. 
First, the two quantities are identical in the Kitaev limit while they can be generally different by non-Kitaev interactions. Second, the two quantities are in the same representation of the $D_3$ group, so $\bar{\nu}({\bf h})$ and $\nu({\bf h})$ have in common the symmetric zeroes. Third, differences between the two quantities may be understood by considering other possible $A_2$ representation spin operators that may contribute to the topological invariant. For example, linear and higher-order spin operators exist in addition to the chiral operator.
Since the topological invariant $\nu$ is related with the thermal Hall conductivity $\kappa_{xy}$, the associated energy current operator directly informs us  of relevant spin operators to the topological invariant.
We find that linear spin operator is irrelevant to $\kappa_{xy}$ and $\nu$ (Supplementary~Note~7).
We also evaluate the expectation values of higher-order spin operators for the KQSL and confirm that their sizes are substantially small compared to the chirality (Supplementary~Note~8). Therefore, we argue that the critical lines of the non-abelian KQSL are mainly determined by the zero lines of the chirality. \\

\section*{Discussion}

Intrinsic topological properties of the non-abelian KQSL including the critical lines, the symmetric zeroes, and the cubic dependence are highlighted in this work by exploiting the symmetries of time reversal and $D_3$ point group. 
The chirality operator is used to evaluate the topological properties of the KQSL via the ED calculations.
Now we discuss how the properties are affected by lattice symmetry breaking such as stacking faults in real materials.
First, the existence of the critical lines relies on time reversal, 
thus it is not destroyed by lattice symmetry breaking.
The symmetric zeroes appearing at the bond directions, however, are a consequence of the $D_3$ symmetry.
The locations of the zeroes get shifted upon breaking the symmetry,
which is confirmed in ED calculations of the chirality.

\begin{figure*}[tb]
\centering
\includegraphics[width=0.7\linewidth]{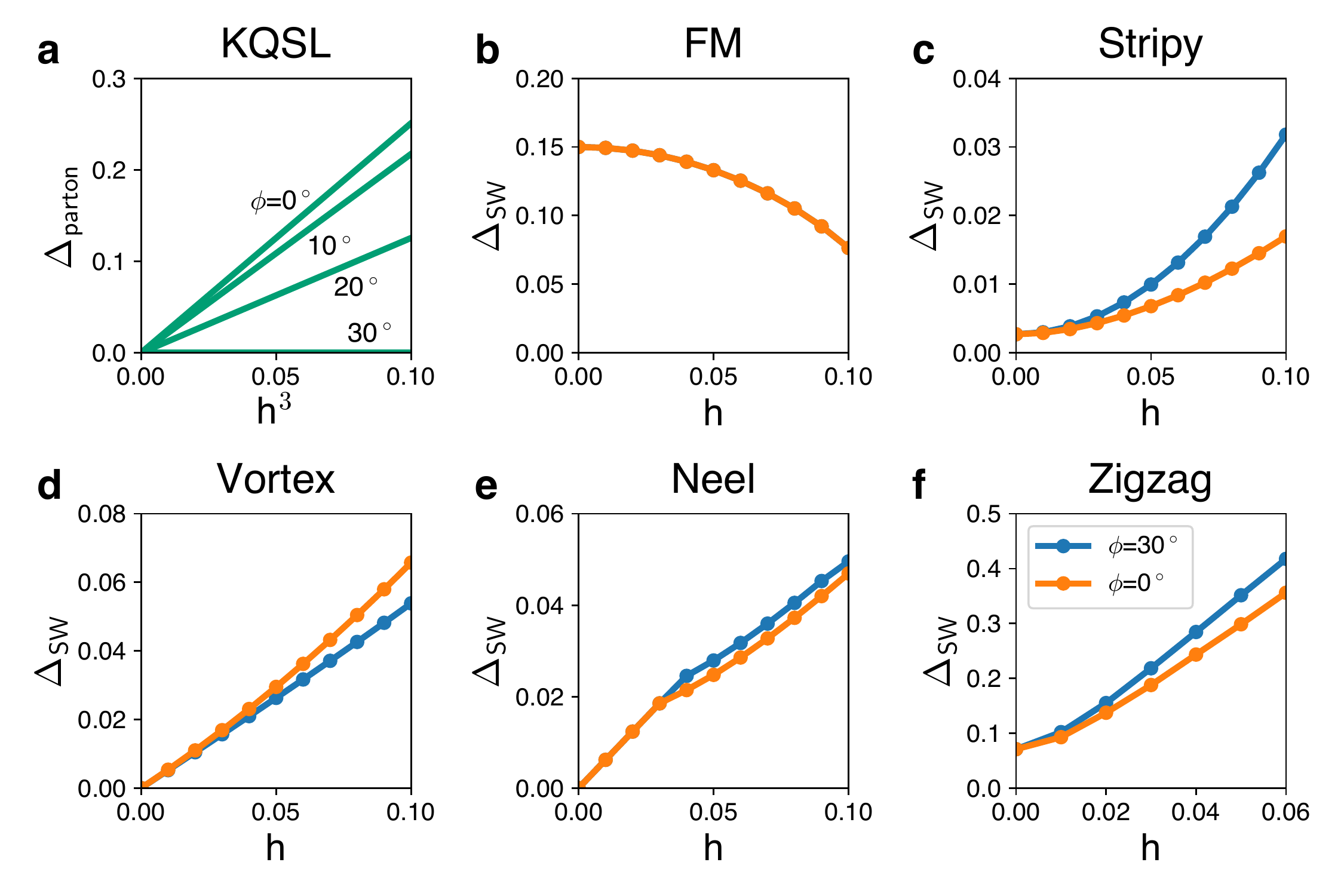}
\caption{{\bf Comparison of the KQSL with magnetically ordered phases.}
{\bf a} KQSL: Majorana energy gap $\Delta_{\rm parton}$ as a function of $h^3$ for the in-plane fields, $(\theta=90^\circ,\phi=0^\circ,10^\circ,20^\circ,30^\circ)$, obtained by a parton theory.
{\bf b}-{\bf f} Ferromagnetic (FM), stripy, vortex, Neel, and zigzag phases: Magnon gap $\Delta_{\rm SW}$ as a function of $h$ for the in-plane fields, $(\theta=90^\circ,\phi=0^\circ,30^\circ)$, obtained with a spin wave theory for the parameter sets $\#5\sim9$ in Table~\ref{tab:1}.
}  
\label{fig:4}
\end{figure*}

The cubic dependence for in-plane fields also originates from the $D_3$ symmetry and topology in the KQSL.
The characteristic nonlinear response is not expected in magnetically ordered phases, which we check by performing spin wave calculations.
Figure~\ref{fig:4} contrasts the KQSL with magnetically ordered phases in terms of excitation energy gap (Majorana gap vs. magnon gap).
The magnetic phases show completely different behaviors from the $h$-cubic dependence.
Hence, the characteristic cubic dependence under in-plane magnetic fields may serve as an experimentally measurable signature of the KQSL.

\begin{figure*}[ht]
\centering
\includegraphics[width=0.7\linewidth]{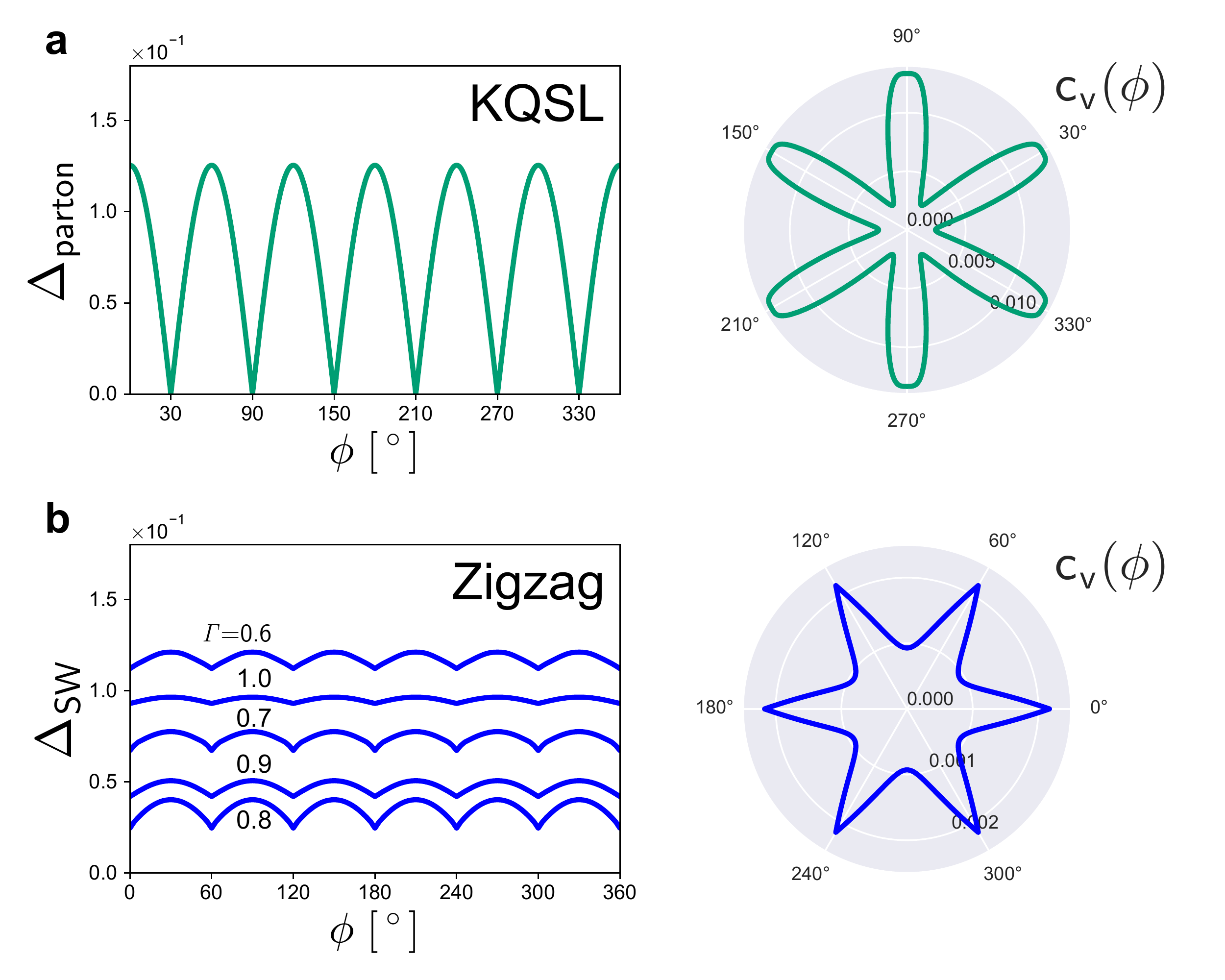}
\caption{{\bf Magnetic field angle dependence of specific heat in the KQSL and zigzag states.}
{\bf a} KQSL state: Majorana gap $\Delta_{\rm parton}$ and specific heat $c_v$ as functions of in-plane field angle $\phi$, obtained by a parton theory.
{\bf b} Zigzag state: magnon gap $\Delta_{\rm SW}$ and specific heat $c_v$ as functions of in-plane field angle $\phi$, obtained by a spin wave theory with $(K,\varGamma,\varGamma',h)=(-1,0.8,-0.05,0.03)~\&~k_BT=0.01$.
Magnon gaps for other values of $\varGamma$ are shown together to highlight the generality of the field angle dependence.
}  
\label{fig:5}
\end{figure*}

The universal properties of the KQSL can be observed by heat capacity experiments.
Figure~\ref{fig:5}a illustrates the calculated specific heat $c_v(\phi)$ for the KQSL as a function of in-plane field angle $\phi$ (where magnetic field is rotated within the honeycomb plane).
The specific heat is maximized by gapless continuum of excitations when the magnetic field is aligned to the bond directions. 
For comparison, the zigzag state, observed in $\alpha$-RuCl$_3$ at zero field, is investigated by using a spin wave theory.
The magnon spectrum is gapped due to completely broken spin rotation symmetry, so there is no critical line on the $(\theta, \phi)$ plane (Fig.~\ref{fig:5}b).
Compared to the KQSL, the zigzag state exhibits reverted patterns of $\phi$ dependence in the excitation energy gap and specific heat.
The energy gap is maximized and the specific heat is minimized at the bond directions.
This behavior is closely related with the structure of spin configuration: all spin moments are aligned perpendicular to a certain bond direction selected by magnetic field direction (Supplementary~Note~9).
The distinct patterns of $\phi$ dependence in Fig.~\ref{fig:5} characterize differences between the non-abelian KQSL and zigzag states.
Remarkably, such behaviors were observed in the recent heat capacity experiments with in-plane magnetic fields~\cite{Shibauchi2020specificheat}.
Covering the polar angle ($\theta$) in the heat capacity measurements will provide more detailed information on the critical lines and spin interactions in $\alpha$-RuCl$_3$ (see Fig.~\ref{fig:1}d). 

Lastly, we have examined the chirality and critical behaviors of excitation energy gap for magnetically ordered phases of $H(\theta,\phi)$.
It is found that the associated magnon gap does not have any critical lines, and there is no resemblance/correlation between the magnon gap and the chirality (Supplementary~Note~9).

To summarize, we have uncovered characteristics of the non-abelian Kitaev quantum spin liquid, including the topologically protected critical lines, the symmetric zeroes, and the $h$-cubic dependence for in-plane fields, by using ED calculations with the chirality operator.
Furthermore, we characterize the topological fingerprints of the KQSL in heat capacity.
We expect our findings to be useful guides for identifying the KQSL in candidate materials such as $\alpha$-RuCl$_3$.
Investigation of the universal properties in field angle dependence of thermodynamic quantities such as spin susceptibility is highly desired, and it would be also useful to apply our results to the recently studied field angle dependence of thermodynamic quantities\cite{Vojta2018,Shibauchi2020specificheat,Modic2018,Riedl2019,Kee2021,Bachus2021}.

\section*{Methods}

\begin{figure*}[t]
\centering
\includegraphics[width=0.85\linewidth]{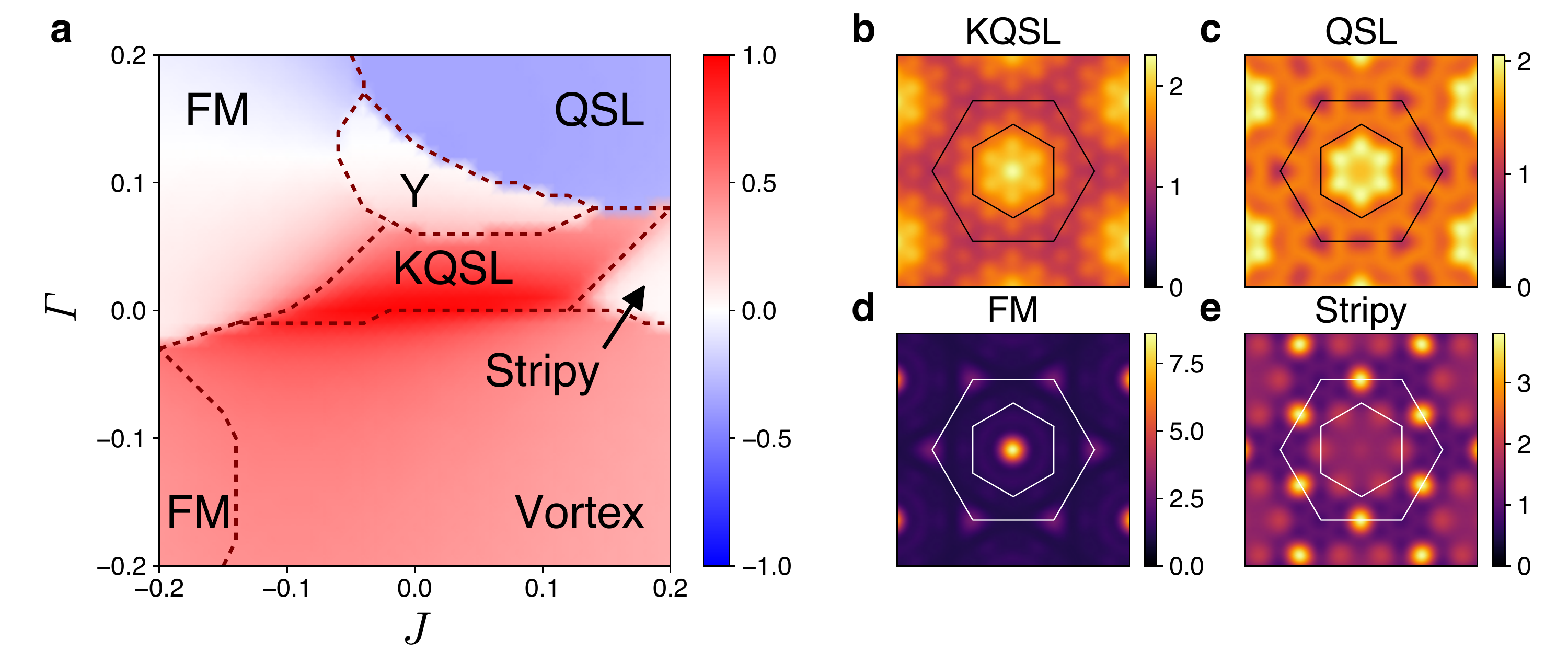}
\caption{{\bf Ferromagnetic KQSL and nearby magnetic states.}
{\bf a} Phase diagram of $H_{KJ\varGamma\varGamma^\prime}$ at $K=-1$ and $\varGamma'=0.05$.
The color encodes the flux operator expectation value $\langle \hat{W}_p \rangle$, and the dashed lines denote phase boundaries determined by the ground state energy second derivatives $-\partial^2 E_{\rm gs}/\partial \xi^2~(\xi=J,\varGamma)$.
{\bf b-e} Color maps of the spin structure factor $S({\bf q})$ for the KQSL, QSL, ferromagnetic (FM), and stripy states.
The inner and outer hexagons denote the first and second Brillouin zones of the honeycomb lattice.
}  
\label{fig:6}
\end{figure*}

\begin{figure*}[t]
\centering
\includegraphics[width=0.85\linewidth]{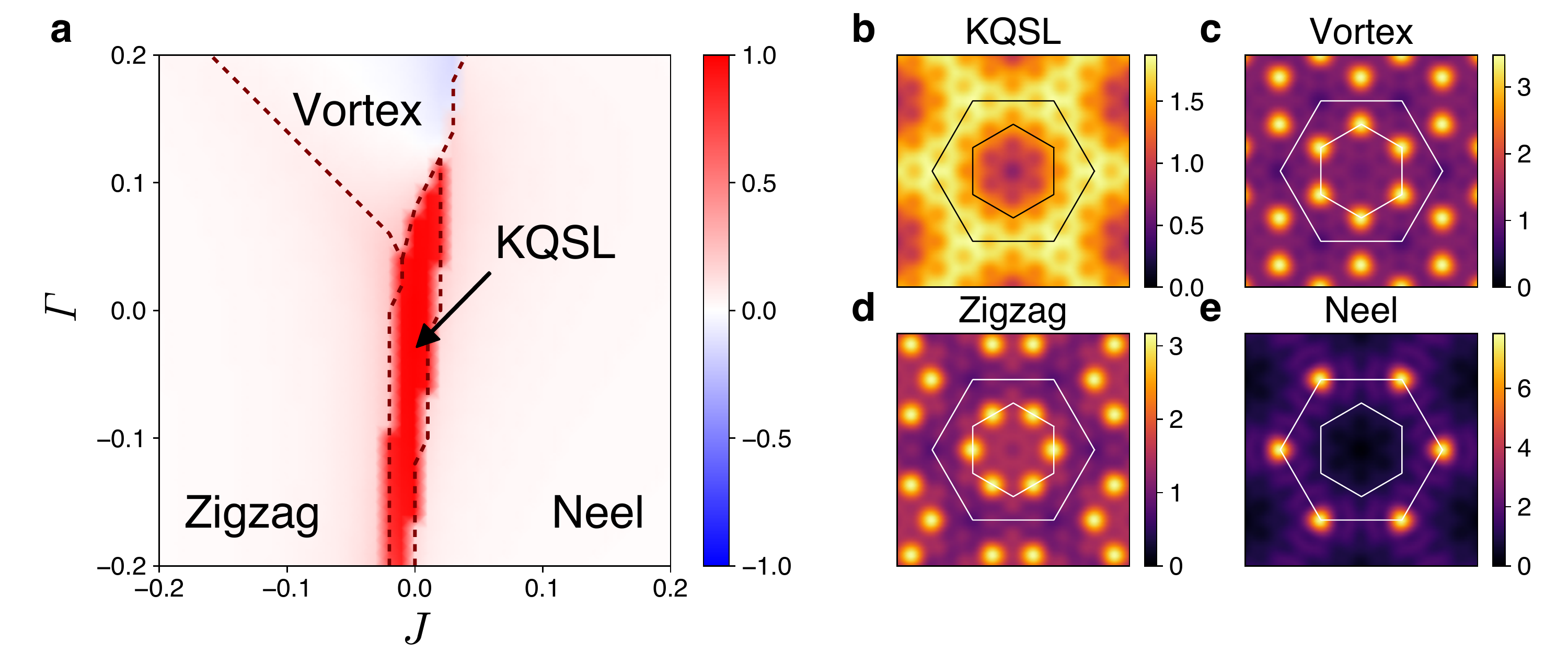} 
\caption{
{\bf Antiferromagnetic KQSL and adjacent magnetic states.}
{\bf a} Phase diagram of $H_{KJ\varGamma\varGamma^\prime}$ at $K=1$ and $\varGamma'=-0.05$.
The color encodes the flux operator expectation value $\langle \hat{W}_p \rangle$, and the dashed lines denote phase boundaries determined by the ground state energy second derivatives $-\partial^2 E_{\rm gs}/\partial \xi^2~(\xi=J,\varGamma)$.
{\bf b-e} Color maps of the spin structure factor $S({\bf q})$ for the KQSL,  vortex, zigzag, and Neel states.
The inner and outer hexagons denote the first and second Brillouin zones of the honeycomb lattice.}  
\label{fig:7}
\end{figure*}

\noindent{\bf Exact diagonalization.} The KQSL and other magnetic phases of $H(\theta,\phi)$ are mapped out 
by using the flux operator $\hat{W}_p = 2^6 S_1^z S_2^y S_3^x S_4^z S_5^y S_6^x $, the second derivative of the ground state energy $\partial^2 E_{\rm gs}/\partial \xi^2~(\xi=J,\varGamma,\varGamma')$, and the spin structure factor
$ S({\bf q})=\frac{1}{N}\sum_{i,j} \langle {\bf S}_i \cdot {\bf S}_j \rangle e^{i{\bf q}\cdot({\bf r}_i-{\bf r}_j)} $.
We find that the KQSL differently responds to non-Kitaev interactions depending on the sign of the Kitaev interaction.
Furthermore,  the non-abelian phase of the KQSL is ensured by checking topological degeneracy and modular $\mathcal{S}$ matrix~\cite{Kitaev,Hickey2019,MES_Zhang2012}.

Figures~\ref{fig:6}~and~\ref{fig:7} display the phase diagrams of $H_{KJ\varGamma\varGamma^\prime}$. 
A different structure of phase diagram is found depending on the sign of the Kitaev interaction.
With the ferromagnetic Kitaev coupling ($K<0$ as in Fig.~\ref{fig:6}), 
the KQSL takes an elongated region along the $J$ axis but substantially narrowed along the $\varGamma$ axis, showing the sensitivity to the $\varGamma$ coupling of the ferromagnetic KQSL.
Crossover-type continuous transitions are mostly observed
among the KQSL 
and nearby magnetically ordered states such as ferromagnetic, stripy and vortex states~\cite{Rau2014,Chaloupka2015,Chaloupka2019}.
Nature of the phase Y is unclarified within the finite size calculation. 
Unlike the aforementioned magnetically ordered states (Fig.~\ref{fig:6}d-e),
the phase Y does not exhibit sharp peaks and periodicity in the spin structure factor, from which the phase is speculated to have an incommensurate spiral order or no magnetic order.
It is remarkable that another quantum spin liquid phase, characterized by negative $\langle \hat{W}_p \rangle$, exists in a broad region of the phase diagram (blue region of Fig.~\ref{fig:6}a)~\cite{Kee2019}.
The QSL and KQSL show similarity in the spin structure factor (Fig.~\ref{fig:6}b-c).
Nonetheless, the QSL as well as the phase Y get suppressed when the sign of $\varGamma'$ is changed to negative.
A zigzag antiferromagnetic order instead sets in under negative sign of $\varGamma'$ (Supplementary~Fig.~6).

In case of the antiferromagnetic Kitaev coupling ($K>0$ as in Fig.~\ref{fig:7}), the KQSL is found to be more sensitive to the Heisenberg coupling rather than the $\varGamma$ coupling, and surrounded by magnetically ordered states such as the vortex, zigzag, and Neel states~\cite{Rau2014,Chaloupka2015,Chaloupka2019}.
In contrast to the ferromagnetic KQSL case, phase transitions between the antiferromagnetic KQSL and adjacent ordered states are all discontinuous as shown by $\langle \hat{W}_p \rangle$ in Fig.~\ref{fig:7}a.
We also find that the antiferromagnetic KQSL and ferromagnetic KQSL are distinguished by different patterns of spin structure factor (Figs.~\ref{fig:6}b~and~\ref{fig:7}b).
Further phase diagrams for other values of $\varGamma'$ are provided in Supplementary~Fig.~6.

We also examine the phase diagrams at weak magnetic fields, and confirm that the overall structures remain the same as the zero-field results.
We find that the chirality is useful for the identification of distinct phase boundaries. In some cases, the chirality performs better than the conventionally used flux (Supplementary~Note~12).

We ensure the non-abelian KQSL phase by checking the Ising anyon topological order via threefold topological degeneracy~\cite{Kitaev,topo_deg_Kells2009} and modular $\mathcal{S}$ matrix~\cite{MES_Zhang2012,Sheng2014,Hickey2019}.
As an example, the $\mathcal{S}$ matrix
\begin{eqnarray}
\mathcal{S}_{\rm ED}&=&\langle \Psi^{\rm MES-I} | \Psi^{\rm MES-II} \rangle
\nonumber\\
&=&
\left[
\begin{array}{ccc}
0.45e^{-i0.08} & 0.53e^{-i0.03} & 0.70e^{i0.04}
\\
0.53e^{-i0.03} & 0.50 & -0.71e^{-i0.01}
\\
0.70e^{i0.04} & -0.71e^{-i0.01} & 0.02e^{-i2.09}
\end{array}
\right]
\nonumber\\
& \approx &
\left[
\begin{array}{ccc}
1/2 & 1/2 & 1/\sqrt{2}
\\
1/2 & 1/2 & -1/\sqrt{2}
\\
1/\sqrt{2} & -1/\sqrt{2} & 0
\end{array}
\right]
\label{eq:modular}
\end{eqnarray}
is obtained for the parameter set~$\# 4$ in Table~\ref{tab:1} with the magnetic field fixed along the [111] direction ($\theta=0^\circ$).
See Supplementary Note 10 for the topological degeneracy and modular matrix computation.
\\

\section*{Data availability}
The data that support the findings of this study are available from the corresponding author on reasonable request.

\section*{Code availability}
The code used to generate the data in this study is available from the corresponding author upon reasonable request.

\bibliographystyle{naturemag}


\section*{Acknowledgements}
We thank L. Balents, J. H. Han, Y. B. Kim, and Y. Matsuda for invaluable discussions. 
We also thank B. H. Kim and H.-Y. Lee for useful discussions at the early stage of the project.
This work was supported by Institute for Basic Science under Grants No.~IBS-R024-D1 (AG),
Korea Institute for Advanced Study under Grant No.~PG071401~\&~PG071402 (KH),
 and NRF of Korea under Grant  NRF-2021R1C1C1010429 (AG), NRF-2019M3E4A1080411, NRF-2020R1A4A3079707, and NRF-2021R1A2C4001847 (EGM). Work in Japan was supported by a Grant-in-Aid for Scientific Research on innovative areas ``Quantum Liquid Crystals'' (JP19H05824) from Japan
Society for the Promotion of Science (JSPS), and by JST CREST (JPMJCR19T5).
We thank Center for Advanced Computation (CAC) at Korea Institute for Advanced Study (KIAS) for providing computing resources for this work.

\newpage
\section*{Author contributions}
E.-G.M. conceived and supervised the project.  
K.H., J.H.S., and A.G. performed theoretical calculations. 
K.H. and T.S. contributed to analysis and comparison of theoretical results and experimental data.  
K.H., A.G., and E.-G.M. prepared the manuscript with the inputs from T.S.

\section*{Competing interests}
The authors declare no competing interests.


\setcounter{equation}{0}
\setcounter{figure}{0}
\setcounter{table}{0}
\setcounter{subsection}{0}

\renewcommand{\thesubsection}{\normalsize Supplementay~Note~\arabic{subsection}}  
\renewcommand{\theequation}{\arabic{equation}}
\renewcommand{\thefigure}{\arabic{figure}}
\renewcommand{\thetable}{\arabic{table}}

\widetext
\pagebreak

\subsection*{\large Supplementary Information on ``Identification of a Kitaev Quantum Spin Liquid by  Magnetic Field Angle Dependence''}

\vspace{1cm}

\subsection{Magnetic field in various coordinate systems}
We use two different coordinate systems defined by $\{\hat{x},\hat{y},\hat{z}\}$ and $\{\hat{a},\hat{b},\hat{c}\}$ for magnetic field:
\begin{equation}
{\bf h}=h_x\hat{x}+h_y\hat{y}+h_z\hat{z}=h_a\hat{a}+h_b\hat{b}+h_c\hat{c}.
\end{equation}
The axes $\{\hat{x}, \hat{y}, \hat{z}\}$ are 
the cubic axes of the octahedra enclosing the honeycomb lattice.
The other axes $\{\hat{a}\equiv\frac{1}{\sqrt{6}}(\hat{x}+ \hat{y}-2\hat{z}), \hat{b}\equiv \frac{1}{\sqrt{2}}(-\hat{x}+\hat{y}), \hat{c}\equiv\frac{1}{\sqrt{3}}(\hat{x}+ \hat{y}+\hat{z}) \}$ correspond to the bond-perpendicular, bond-parallel, and out-of-plane axes of the honeycomb lattice (Fig.~1a).
In the spherical coordinate system ($h,\theta,\phi$), the magnetic field is written as
\begin{equation}
h_a=h \sin\theta\cos\phi,~h_b=h\sin\theta\sin\phi,~h_c=h\cos\theta,
\end{equation}
and
\begin{eqnarray}
h_x &=& h \left( -\frac{1}{\sqrt{2}} \sin\theta \sin\phi + \frac{1}{\sqrt{6}} \sin\theta \cos\phi + \frac{1}{\sqrt{3}} \cos\theta \right),~~~
\\
h_y &=& h \left( \frac{1}{\sqrt{2}} \sin\theta \sin\phi + \frac{1}{\sqrt{6}} \sin\theta \cos\phi + \frac{1}{\sqrt{3}} \cos\theta \right),~~~
\\
h_z &=& h \left( -\sqrt{\frac{2}{3}} \sin\theta \cos\phi + \frac{1}{\sqrt{3}} \cos\theta \right).~~~
\end{eqnarray} 
The polar and azimuthal angles ($\theta,\phi$) are measured from the out-of-plane axis $\hat{c}$ and the bond-perpendicular axis $\hat{a}$, respectively (Fig.~1b).

\begin{table}[b]
\begin{ruledtabular}
\begin{tabular}{ll}
Operation & Field-angle transformation
\\
\hline
Time reversal & $H(\theta,\phi) \rightarrow H(180^{\circ}-\theta,180^{\circ}+\phi)$
\\
Spatial Inversion & $H(\theta,\phi) \rightarrow H(\theta,\phi)$
\\
$C_3$ rotation & $H(\theta,\phi) \rightarrow H(\theta,120^{\circ}+\phi)$
\\
$C_2$ rotation (about $z$-bond)  & $H(\theta,\phi) \rightarrow H(180^{\circ}-\theta,180^{\circ}-\phi)$
\end{tabular}
\end{ruledtabular}
\renewcommand{\tablename}{\textbf{Supplementary Table}}
\caption{\bf Symmetry operations of $H_{JK\varGamma\varGamma'}$ and the associated field-angle transformation rules on $H(\theta,\phi)$. 
}
\label{tab:symm}
\end{table}

\subsection{Symmetry}

The symmetries of $H_{JK\varGamma\varGamma'}$ are as follows.
\begin{itemize}
\item {\it Time reversal} : ${\bf S}_i \rightarrow -{\bf S}_i$.
\item {\it Spatial inversion} about each bond center of the honeycomb lattice : ${\bf S}_i \rightarrow {\bf S}_{i'}$.
\item {\it $C_3$ rotation} about the normal axis to each hexagon plaquette : $ (S_i^x,S_i^y,S_i^z) \rightarrow (S_{i'}^z,S_{i'}^x,S_{i'}^y)$.
\item {\it $C_2$ rotation} about each bond axis : $ (S_i^x,S_i^y,S_i^z) \rightarrow (-S_{i'}^y,-S_{i'}^x,-S_{i'}^z)$ for $z$-bond rotations.
\end{itemize}
See Fig.~1a for visualizations of the $C_3$ and $C_2$ rotation axes.
Transformation rules of $H(\theta,\phi)$ under the symmetry operations are provided in Supplementary~Table~\ref{tab:symm}.

\subsection{Coupling between the chirality operator and magnetic field}

The full expression of the chirality operator $\chi_p$ is given by
\begin{equation}
\hat{\chi}_p
=
S_2^xS_1^zS_6^y + S_5^xS_4^zS_3^y
+S_6^zS_5^yS_4^x + S_3^zS_2^yS_1^x
+S_4^yS_3^xS_2^z + S_1^yS_6^xS_5^z
\end{equation}
with the site convention shown in Fig. 1a.

We determine the most generic coupling form between the chirality operator and magnetic field using symmetry.
Suppose we have $f({\bf h}) S_2^xS_1^zS_6^y$ at neighboring three sites 2-1-6 in Fig.~1a. Then $f({\bf h})$ should be a polynomial consisting of only odd powers of $h_{x,y,z}$ by time reversal.
$C_2$ rotation (about the $z$-bond axis passing through the site 1) further constrains $f({\bf h})$ into the form
\begin{eqnarray}
&&
f_{xy,z}({\bf h})
\nonumber\\
&&
=
\lambda_1 (h_x + h_y) + \lambda_2 h_z + \lambda_3 h_x h_y h_z
\nonumber\\
&&
+ 
\lambda_4 (h_x^3 + h_y^3) + \lambda_5 h_z^3
+
\lambda_6 ( h_x^2 h_y + h_x h_y^2 ) 
\nonumber\\
&&
+ 
\lambda_7 ( h_x^2 h_z + h_y^2 h_z ) 
+
\lambda_8 ( h_x h_z^2 + h_y h_z^2 )
+
\mathcal{O}(h^5).
\label{Seq:f}
\end{eqnarray}
Lastly, we apply $C_3$ rotation and spatial inversion to $f_{xy,z}({\bf h}) S_2^xS_1^zS_6^y$, which generates symmetry-related other five terms at a hexagon plaquette:
\begin{equation}
f_{xy,z}({\bf h}) ( S_2^xS_1^zS_6^y + S_5^xS_4^zS_3^y )
+
f_{zx,y}({\bf h}) ( S_6^zS_5^yS_4^x + S_3^zS_2^yS_1^x )
+
f_{yz,x}({\bf h}) ( S_4^yS_3^xS_2^z + S_1^yS_6^xS_5^z ),
\label{Seq:chi}
\end{equation}
where $f_{yz,x}$ and $f_{zx,y}$ are obtained by cyclic permutations of $h_{x,y,z}$ in $f_{xy,z}$.
Summing  the contributions from all hexagon plaquettes, we arrive at the final expression
\begin{equation}
\sum_{\langle ij \rangle_{\alpha}\langle jk \rangle_{\beta}}
f_{\alpha\beta,\gamma}({\bf h})
 S_i^{\alpha} S_j^{\gamma} S_k^{\beta}.
 \label{eq:chi-h-coupling}
\end{equation}
These effective interactions are generated by applied magnetic fields, and the coupling functions $f_{\alpha\beta,\gamma}({\bf h})$ can be evaluated by the perturbation theory shown below.

Our major finding is summarized before the construction of the perturbation theory.
The coupling functions $f_{\alpha\beta,\gamma}({\bf h})$ determine topological properties of the non-abelian Kitaev quantum spin liquid (KQSL).
In particular, the chirality is given by
\begin{equation}
\chi = \frac{1}{N} \sum_p \langle \hat{\chi}_p \rangle \approx -[f_{xy,z}({\bf h})+f_{yz,x}({\bf h})+f_{zx,y}({\bf h})],
\end{equation}
and its sign and magnitude are closely related with the Chern number and the energy gap of  Majorana fermion excitations.
The chirality can be arranged in terms of $A_2$ representations of ${\bf h}$:
\begin{equation}
\chi\approx\Lambda_1F_1+\Lambda_3F_3+\Lambda'_3F'_3+\Lambda''_3F''_3+\mathcal{O}(h^5),
\end{equation}
where the coefficients ($\Lambda$) and the $A_2$ representations ($F$) are listed in Supplementary~Table~\ref{tab:A2-expansion}.
The $\Lambda$ coefficients are obtained by the perturbation theory below.

\begin{table*}[b]
\begin{ruledtabular}
\begin{tabular}{ll}
Coefficient & $A_2$ representation
\\
\hline
$\Lambda_1=-(2\lambda_1+\lambda_2)$ & $F_1({\bf h})=h_x+h_y+h_z$
\\
$\Lambda_3=-3\lambda_3$ & $F_3({\bf h})=h_xh_yh_z$
\\
$\Lambda'_3=-(2\lambda_4+\lambda_5)$ & $F'_3({\bf h})=h_x^3+h_y^3+h_z^3$
\\
$\Lambda''_3=-(\lambda_6+\lambda_7+\lambda_8)$ & $F''_3({\bf h})=h_x^2h_y+h_xh_y^2+h_y^2h_z+h_yh_z^2+h_z^2h_x+h_zh_x^2$
\end{tabular}
\end{ruledtabular}
\renewcommand{\tablename}{\textbf{Supplementary Table}}
\caption{\bf Expansion of the chirality $\chi$ in terms of $A_2$ representations of ${\bf h}$. 
}
\label{tab:A2-expansion}
\end{table*}

\subsection{Perturbative expansion}

The perturbation theory is developed by decomposing the full Hamiltonian $H$ into the unperturbed part ($H_0=\sum_{\langle jk \rangle_{\gamma}} K S_j^{\gamma} S_k^{\gamma}$) and perturbation part ($H'=H-H_0$).
In this setup, we focus on the zero-flux sector of the pure Kitaev model $H_0$ ({\it i.e.,} $\langle \hat{W}_p \rangle=1$). 
 
For a systematic derivation, we employ a quasi-degenerate perturbation theory~\cite{sref_Winkler2003} and construct an effective Hamiltonian $\mathcal{H}$ in the zero-flux sector.
Up to the third order of $H'$, the effective Hamiltonian $\mathcal{H}$ is formally given by
\begin{equation}
\mathcal{H} 
=
PH_0P
-
\frac{1}{\Delta_{\textup{flux}}} P H' Q H' P
+
\frac{1}{\Delta_{\textup{flux}}^2} P H' Q H' Q H' P
-
\frac{1}{2\Delta_{\textup{flux}}^2} (P H' P H' Q H' P + {\rm H.c.} ),
\label{Seq:perturbation}
\end{equation}
where $P$ is the projection operator into the zero-flux sector, and $Q=1-P$.
Here we have assumed that energy differences between the zero-flux sector and nonzero-flux sectors are simply given by the flux gap $\Delta_{\textup{flux}}(=0.26S^2|K|)=0.065|K|$~\cite{sref_Kitaev}.
In order for the perturbation theory to be valid, non-Kitaev couplings should be much smaller than the flux gap ($J,\varGamma,\varGamma',h \ll\Delta_{\textup{flux}}$).

The $\lambda$ coefficients in Supplementary Eq. (\ref{Seq:f}) are obtained by the perturbative calculations of Supplementary Eq. (\ref{Seq:perturbation}), which is demonstrated by  using the second order term $-\frac{1}{\Delta_{\textup{flux}}} P H' Q H' P$ as an example.
Among various interactions generated by the perturbation, we collect components of the chirality operator, $S_i^{\alpha} S_j^{\gamma} S_k^{\beta}$, which are defined over three neighboring sites $ijk$ connected by two bonds $\langle ij \rangle_{\alpha}$ and $\langle jk \rangle_{\beta}$ with all different $\alpha,\beta,\gamma$.
This type of terms are produced by $P H' Q H' P$ via two processes:
(i) $S_i^{\alpha} S_j^{\gamma}$ from the $\varGamma'$ coupling and $S_k^{\beta}$ from the Zeeman coupling, 
or (ii) $S_i^{\alpha} $ from the Zeeman coupling and $S_j^{\gamma} S_k^{\beta}$ from the $\varGamma'$ coupling.
These second order processes result in the effective interaction
\begin{equation}
-\frac{1}{\Delta_{\textup{flux}}} P H' Q H' P
=
P \left[ \sum_{\langle ij \rangle_{\alpha}\langle jk \rangle_{\beta}}
\frac{2\varGamma'(h_{\alpha}+h_{\beta})}{\Delta_{\textup{flux}}} S_i^{\alpha} S_j^{\gamma} S_k^{\beta} \right] P
+ \cdots 
\label{Seq:Heff_2nd}
\end{equation}
that corresponds to the $\lambda_1$ term of Supplementary Eq. (\ref{Seq:f}).
Note that the components of the chirality operator appear in the second order because of the presence of $\varGamma'$ in drastic contrast to the pure Kitaev model where the chirality operator is generated in the third order~\cite{sref_Takikawa2019}. 

Repeating the perturbative expansion up to the third order, we obtain the effective Hamiltonian
\begin{equation}
\mathcal{H}
=
P\left[
H_0
+
\sum_{\langle ij \rangle_{\alpha}\langle jk \rangle_{\beta}}
f_{\alpha\beta,\gamma}({\bf h})
 S_i^{\alpha} S_j^{\gamma} S_k^{\beta}
\right]P 
+ 
\cdots ,
\label{eq:Heff}
\end{equation}
where the $f$ function is given by Supplementary~Eq.~(\ref{Seq:f}) with the $\lambda$ coefficients:
\begin{eqnarray}
&& \lambda_1 
= 
\frac{2\varGamma'}{\Delta_{\textup{flux}}}
-
\frac{5J\varGamma'}{2\Delta_{\textup{flux}}^2}
+
\frac{2\varGamma\varGamma'}{\Delta_{\textup{flux}}^2}
-
\frac{\varGamma'^2}{2\Delta_{\textup{flux}}^2},
\label{eq:lambda1}
\\
&& \lambda_2 
= 
-
\frac{J\varGamma'}{\Delta_{\textup{flux}}^2}
-
\frac{3\varGamma'^2}{2\Delta_{\textup{flux}}^2},
\label{eq:lambda2}
\\
&& \lambda_3 = -\frac{6}{\Delta_{\textup{flux}}^2}.
\label{eq:lambda3}
\end{eqnarray}
The other coefficients $\lambda_{4,5,6,7,8}$ remain zero ($\lambda_{4,5,6,7,8}=0$) up to third order.

\subsection{Parton theory}

\begin{figure}[tb]
\centering
\includegraphics[width=0.6\linewidth]{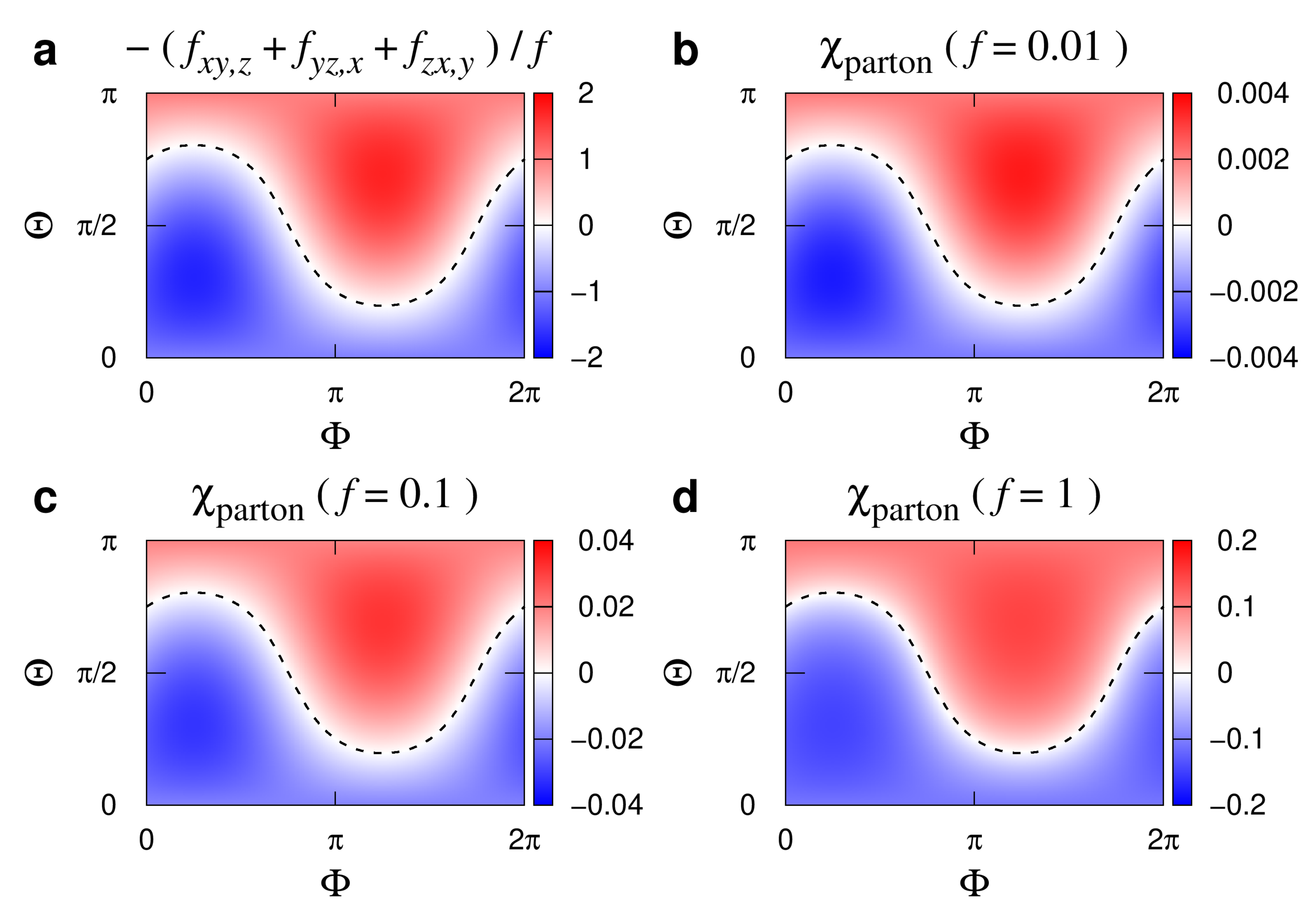} 
\renewcommand{\figurename}{\textbf{Supplementary Figure}}
\caption{{\bf Chirality $\boldsymbol{\chi_{\rm parton}}$ of the parton theory.}
{\bf a-d} Color maps of $-
(f_{xy,z}+f_{yz,x}+f_{zx,y})/f$ and $\chi_{\rm parton}$ ($f=0.01,~0.1,~1$) as functions of $\Theta$ and $\Phi$.
In each plot, the dashed line indicates the points of $f_{xy,z}+f_{yz,x}+f_{zx,y}=0$.
}
\label{fig:S1}
\end{figure}

\noindent{\bf Energy gap.}
The component of the chirality operator $S_i^{\alpha} S_j^{\gamma} S_k^{\beta}$ gives rise to a next-nearest-neighbor hopping term of $c$-Majorana fermions without changing the flux sector~\cite{sref_Kitaev,sref_Baskaran2007}.
This can be seen from the identity
\begin{equation}
S_i^{\alpha} S_j^{\gamma} S_k^{\beta}
= -\frac{i}{8} D_j \hat{u}_{ij} \hat{u}_{kj} c_i c_k ,
\label{eq:SSS}
\end{equation}
where $\hat{u}_{ij}=ib_i^{\alpha}b_j^{\alpha}$, $\hat{u}_{kj}=ib_k^{\beta}b_j^{\beta}$, and $D_j = b_j^x b_j^y b_j^z c_j$.
Applying this to $\mathcal{H}$ in Supplementary Eq. (\ref{eq:Heff}), we obtain the effective Majorana Hamiltonian
\begin{small}
\begin{equation}
\mathcal{H}
=
P
\left[
\sum_{\langle jk \rangle_{\gamma}} \frac{-iK}{4} \hat{u}_{jk} c_j c_k
+
\sum_{\langle ij \rangle_{\alpha}\langle jk \rangle_{\beta}}
\frac{-if_{\alpha\beta,\gamma}}{8} D_j \hat{u}_{ij} \hat{u}_{kj} c_i c_k
\right]P .
\end{equation}
\end{small}For the calculations of fermion excitation spectrum,
we choose the uniform configuration of $\mathbb{Z}_2$ gauge fields: $\hat{u}_{jk}=1$ for $j\in A$, $k\in B$ (or $\hat{u}_{jk}=-1$ for $j\in B$, $k\in A$), where $A,B$ imply two sublattices of the honeycomb lattice.
After the gauge-fixing and Fourier transformation to momentum space, the Hamiltonian is written as
\begin{equation}
\mathcal{H}=\frac{1}{4}\sum_{\bf q} 
\left[
\begin{array}{cc}
c_{-{\bf q}}^A & c_{-{\bf q}}^B
\end{array}
\right]
\left[
\begin{array}{cc}
M_{\bf q} & i U_{\bf q}
\\
-i U_{\bf q}^* & -M_{\bf q}
\end{array}
\right]
\left[
\begin{array}{c}
c_{{\bf q}}^A
\\
c_{{\bf q}}^B
\end{array}
\right],
\end{equation}
where the subscript ${\bf q}$ denotes momentum, and the superscripts $A,B$ imply two sublattices of the honeycomb lattice.
The matrix elements are given by
$
M_{\bf q}
=
-\frac{1}{2} 
[ f_{xy,z} \sin{\bf q}\cdot({\bf n}_2-{\bf n}_1)
+ f_{yz,x} \sin{\bf q}\cdot(-{\bf n}_2)
+ f_{zx,y} \sin{\bf q}\cdot{\bf n}_1
],
$
and
$
U_{\bf q}
=
-\frac{K}{2} (1+e^{i{\bf q}\cdot{\bf n}_1}+e^{i{\bf q}\cdot{\bf n}_2}),
$
where ${\bf n}_1,~{\bf n}_2$ imply lattice vectors of the honeycomb lattice.
In terms of the complex fermion operators  
$
\Psi_{\bf q}
=
\left[
\psi_{\bf q}
,
\psi_{-{\bf q}}^{\dagger}
\right]^T
\equiv
\frac{1}{2}
\left[
c_{\bf q}^A+ic_{\bf q}^B
,
c_{\bf q}^A-ic_{\bf q}^B
\right]^T
$,
the Hamiltonian takes the form
\begin{equation}
\mathcal{H}=\frac{1}{2} \sum_{\bf q} 
\Psi_{\bf q}^\dagger
~
{\bf R}({\bf q})\cdot\boldsymbol{\sigma}
~
\Psi_{\bf q},
\label{eq:Heff-psi}
\end{equation}
where ${\bf R}({\bf q})=[M_{\bf q},{\rm Im} U_{\bf q},{\rm Re} U_{\bf q}]$ and $\boldsymbol{\sigma}=[\sigma_x,\sigma_y,\sigma_z]$ are the Pauli matrices.
Taking a canonical transformation to the quasiparticle basis $\gamma_{\bf q}=u_{\bf q}\psi_{\bf q}+v_{\bf q}\psi_{-{\bf q}}^\dagger$, we obtain the diagonalized Hamiltonian
$
\mathcal{H}=
E_{\rm gs}
+
\sum_{\bf q} \omega_{\bf q} \gamma_{\bf q}^{\dagger}\gamma_{\bf q}
$
with the excitation spectrum $\omega_{\bf q}= \sqrt{|M_{\bf q}|^2+|U_{\bf q}|^2}$
and the ground state energy $E_{\rm gs}=-\frac{1}{2}\sum_{\bf q}\omega_{\bf q}$.
The gapless points $\pm{\bf q}_*=\pm(\frac{1}{3}{\bf q}_1+\frac{2}{3}{\bf q}_2)$ of the pure Kitaev model are gapped out by the energy gap
\begin{eqnarray}
\Delta_{\rm parton}
=
\omega_{{\bf q}_*}
&=&
\frac{\sqrt{3}}{4}| -(f_{xy,z}+f_{yz,x}+f_{zx,y}) |,~~~
\end{eqnarray}
where ${\bf q}_{1,2}$ are the reciprocal lattice vectors dual to ${\bf n}_{1,2}$.

\noindent{\bf Chern number.}
The gapped spin liquid state is topologically characterized by the Chern number~\cite{sref_Kitaev}
\begin{equation}
\nu_{\rm parton}
=
\int\frac{d^2{\bf q}}{4\pi} [ \hat{\bf R} \cdot {\partial_{q_x} \hat{\bf R}} \times {\partial_{q_y} \hat{\bf R}} ]
=
{\rm sgn} [ -(f_{xy,z}+f_{yz,x}+f_{zx,y})]
\end{equation}
which takes either $+1$ or $-1$ depending on the field direction.

\noindent{\bf Chirality.}
Finally, we compute the expectation value of the chirality operator $\hat{\chi}$:
\begin{equation}
{\chi}_{\rm parton}
=
\frac{1}{N}
\sum_p \langle \hat{\chi}_p \rangle
=
\frac{1}{N}
\left(
\frac{\partial}{\partial f_{xy,z}}
+
\frac{\partial}{\partial f_{yz,x}}
+
\frac{\partial}{\partial f_{zx,y}}
\right) 
E_{\rm gs}
=
-\frac{1}{2N}\sum_{\bf q}
\frac{M_{\bf q}L_{\bf q}}{\omega_{\bf q}},
\end{equation}
where $N$ is the total number of unit cells, and 
$
L_{\bf q}
=
-\frac{1}{2} 
[ \sin{\bf q}\cdot({\bf n}_2-{\bf n}_1)
+ \sin{\bf q}\cdot(-{\bf n}_2)
+ \sin{\bf q}\cdot{\bf n}_1
].
$
By conducting numerical computations for ${\chi}_{\rm parton}$, we find the following property:
\begin{equation}
\chi_{\rm parton}
=
-
(f_{xy,z}+f_{yz,x}+f_{zx,y})~
g,
\end{equation}
where $g$ is a positive function of $(f_{xy,z},f_{yz,x},f_{zx,y})$.
In the limit of $f=\sqrt{f_{xy,z}^2+f_{yz,x}^2+f_{zx,y}} \rightarrow 0$, we find that $g \rightarrow 0.21$.
Supplementary~Fig.~\ref{fig:S1} illustrates the chirality $\chi_{\rm parton}$ with the parametrization
\begin{eqnarray}
f_{yz,x}&=&f\sin\Theta\cos\Phi,
\\
f_{zx,y}&=&f\sin\Theta\sin\Phi,
\\
f_{xy,z}&=&f\cos\Theta,
\end{eqnarray}
where $f>0$, $\Theta\in[0,\pi]$, and $\Phi\in[0,2\pi]$.
On the plane of $\Theta$ and $\Phi$, $\chi_{\rm parton}$ exhibits exactly the same pattern of sign structure as that of $-
(f_{xy,z}+f_{yz,x}+f_{zx,y})$; compare Supplementary~Fig.~\ref{fig:S1}b-d with a.
Therefore, we obtain the relationship
\begin{eqnarray}
\chi_{\rm parton}
&\propto&
-(f_{xy,z}+f_{yz,x}+f_{zx,y}).
\end{eqnarray}
Notice that the chirality itself is intimately related with the Chern number and the Majorana energy gap: $\nu=\textup{sgn}(\chi)$ and $\Delta\sim|\chi|$.

Combining with the results of Supplementary Eqs. (\ref{eq:lambda1})-(\ref{eq:lambda3}),
we obtain the chirality,
\begin{equation}
\chi_{\rm  parton}(\mathbf{h}) \propto \Lambda_1 (h_x +h_y+h_z) + \Lambda_3 h_x h_y h_z,
\label{eq:chirality}
\end{equation}
the Chern number, 
\begin{equation}
\nu_{\rm  parton}({\bf h})
=
{\rm  sgn}[\Lambda_1 (h_x +h_y+h_z) + \Lambda_3 h_x h_y h_z] ,
\label{eq:chern_number}
\end{equation}
and the Majorana energy gap,
\begin{equation}
\Delta_{\rm  parton}(\mathbf{h}) \propto \left| \Lambda_1 (h_x +h_y+h_z) + \Lambda_3 h_x h_y h_z \right|,
\label{eq:parton_gap}
\end{equation}
where $\Lambda_1=-(2\lambda_1+\lambda_2)$ and $\Lambda_3=-3\lambda_3$.
These results are pictorialized in Supplementary~Fig.~\ref{fig:S2}.

\begin{figure*}[t]
\centering
\includegraphics[width=\linewidth]{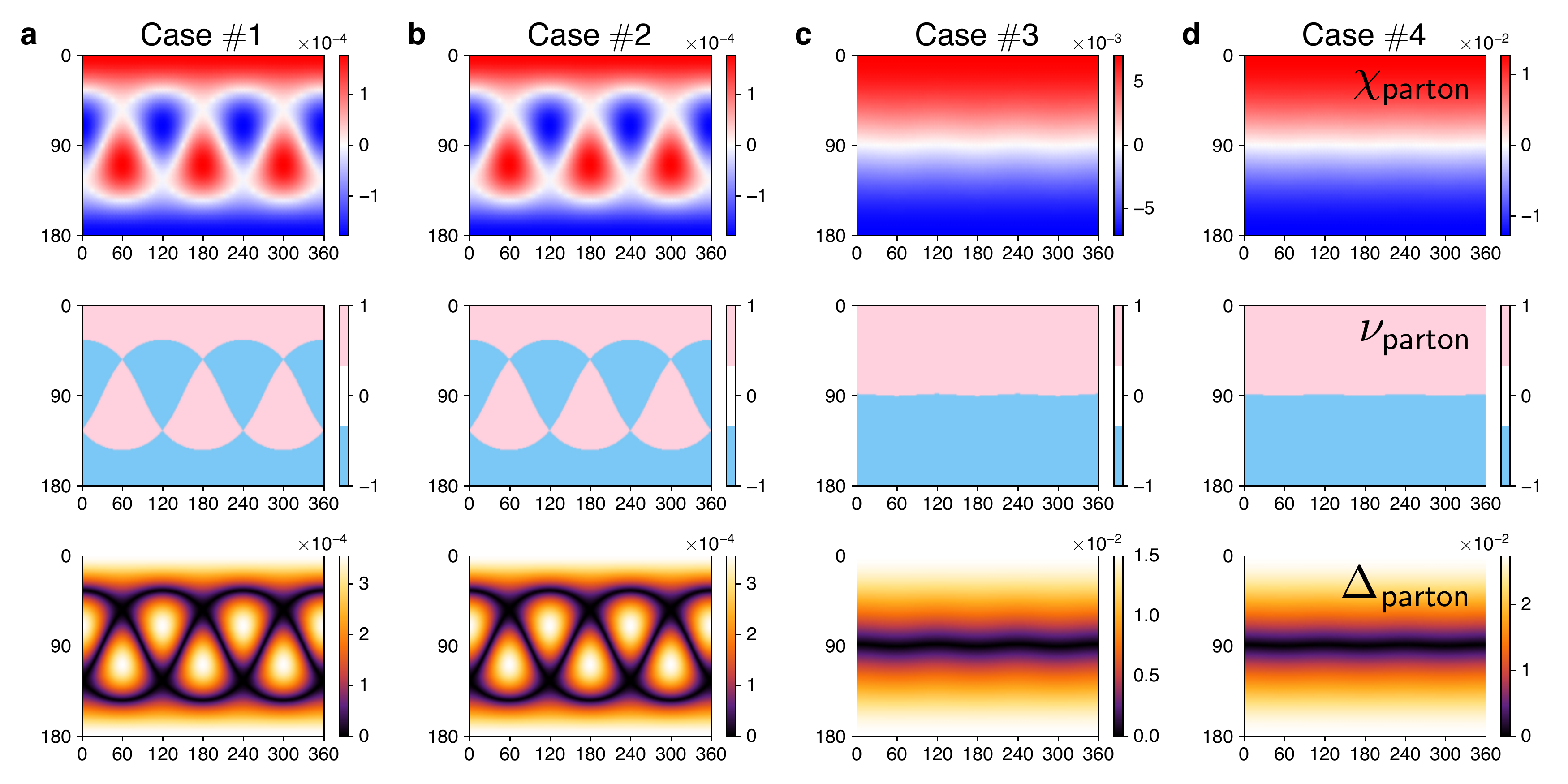} 
\renewcommand{\figurename}{\textbf{Supplementary Figure}}
\caption{{\bf Results of the perturbative parton theory.}
Top, middle, bottom: color maps of the chirality $\chi_{\rm parton}({\bf h})$, the Chern number $\nu_{\rm parton}({\bf h})$, and the Majorana energy gap $\Delta_{\rm parton}({\bf h})$ on the plane of the field angles $(\theta,\phi)$, where the magnetic field strength is fixed by $h=0.01$ (horizontal axis: $\phi~[^\circ]$, vertical axis: $\theta~[^\circ]$).
The parameter sets used in the four cases ($\#1\sim4$) are listed in Table 1.
}
\label{fig:S2}
\end{figure*}

\subsection{Relation between the chirality and Majorana gap}

The relation, $|\chi| \propto \Delta$, between the chirality and the Majorana energy gap has been established by using the perturbative parton analysis near the pure Kitaev model [Supplementary Eqs. (\ref{eq:chirality},\ref{eq:parton_gap})].  
Here, we provide our reasoning why the relation is expected to hold more generally in our KQSL phase diagrams.

First, we find that the relation $|\chi| \propto \Delta$ holds near the universal zeroes for a {\it generic} KQSL with $D_3$ symmetry.
To show this, let us recall that magnetic fields along the bond directions do not break the bond direction $C_2$ symmetry, which makes both $\chi$ and $\Delta$ to be zero.
Suppose we slightly tilt the magnetic field from a bond direction by a small angle $\delta \phi~( \ll 1)$.
The tilting effects appear in the parton Hamiltonian as  
\begin{equation}
\left[
\begin{array}{cc}
0 & iU_{\bf q}
\\
-iU_{\bf q}^* & 0
\end{array}
\right]
\xrightarrow[]{\text{Tilting the field by}~\delta\phi}
\left[
\begin{array}{cc}
m\delta\phi & iU_{\bf q}
\\
-iU_{\bf q}^* & -m\delta\phi
\end{array}
\right].
\end{equation}
It is obvious that Majorana fermions acquire a mass term ($m \delta \phi$), which is an $A_2$ representation of the $D_3$ symmetry.
Then, both of $\chi$ and $\Delta$ are proportional to $\delta\phi$ because they are the same $A_2$ representation, and the linear relationship, $|\chi| \propto \Delta$, is thus established near the universal zeroes. 
We stress that the $D_3$ symmetry plays a significant role in our results. 
In other words, the universal zeroes always appear in a generic KQSL with the $D_3$ symmetry, and the relationship must hold in any points of KQSL in our phase diagrams.

\begin{figure}[tb]
\centering
\includegraphics[width=0.5\linewidth]{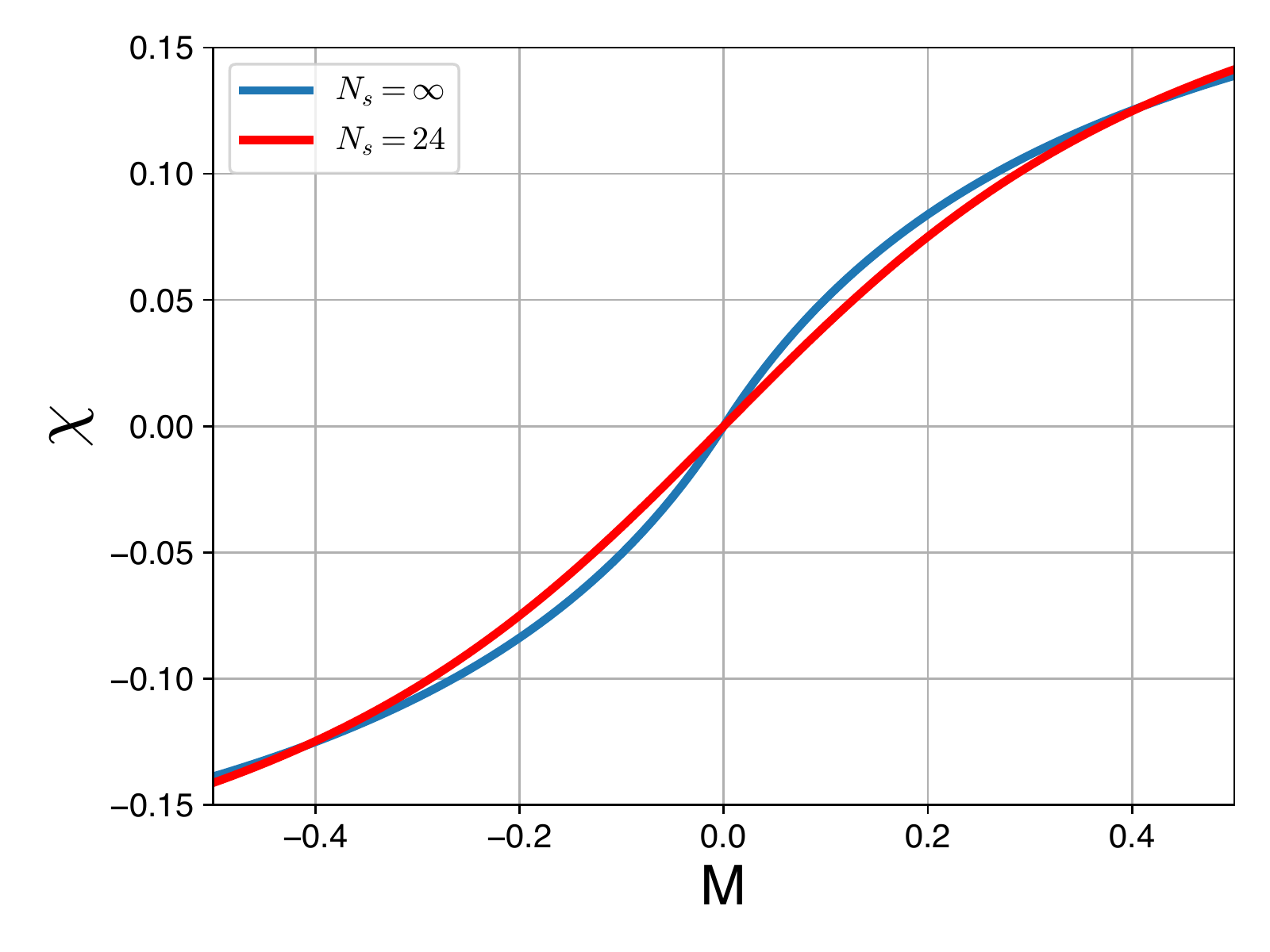}
\renewcommand{\figurename}{\textbf{Supplementary Figure}}
\caption{
{\bf Chirality of the model $H_{\rm \chi}$.}
The chirality $\chi$ as a function of the coupling constant $M$.
Red: ED calculation results for the 24-site cluster.
Blue: parton analysis results for the infinite system.
In all these results, the Kitaev coupling is fixed by $K=-1$.
}
\label{fig:S3}
\end{figure}

Furthermore, we consider another exactly solvable model, 
\begin{equation}
H_{\rm \chi} = K \sum_{\langle jk \rangle_\gamma} S_j^\gamma S_k^\gamma - M \sum_p \hat{\chi}_p ,
\label{eq:R1}
\end{equation}
which consists of the Kitaev interactions ($K$) and the chirality operator three-spin interactions ($M$).
The parton analysis shows that the Chern number is $\nu={\rm sgn}(M)$ and the Majorana gap is $\Delta=\frac{3\sqrt{3}}{4}|M|$.
Supplementary~Fig.~\ref{fig:S3} presents the calculated chirality $\chi$ as a function of $M$; overall, $\chi$ is proportional to $M$. 
The positive correlation between $\chi$ and $M$ is demonstrated, establishing the connection between the chirality and the Marjorana gap in a fairly large window ($|M/K| \lesssim 0.5$).

\subsection{Energy current operator}

To derive the energy current operator ${\bf J}^{\rm E}$, we arrange the Hamiltonian [Eq. (1)] as
\begin{equation}
H=\sum_{\langle jk \rangle} H_{jk}^{\rm EX} + \sum_i H_{i}^{\rm Z},
\end{equation}
where $H_{jk}^{\rm EX}$ is the exchange interaction at bond ${\langle jk \rangle}$, and $H_{i}^{\rm Z}$ is the Zeeman coupling at site $i$.
Then the energy polarization operator can be written as
\begin{equation}
{\bf P}^{\rm E}=\sum_{\langle jk \rangle} \frac{{\bf r}_j+{\bf r}_k}{2} H_{jk}^{\rm EX} + \sum_i {\bf r}_i H_{i}^{\rm Z},
\end{equation}
where ${\bf r}_i$ means the position vector of site $i$.
The energy current is nothing but the time derivative of the polarization~\cite{sref_Han2017,sref_Motome2017}:
\begin{eqnarray}
&&{\bf J}^{\rm E}=\frac{d{\bf P}^{\rm E}}{dt}
=i[H,{\bf P}^{\rm E}]
\\
&=&i\sum_{\langle jk \rangle} \sum_{\langle j'k' \rangle} \frac{{\bf r}_{j'}+{\bf r}_{k'}}{2} [H_{jk}^{\rm EX},H_{j'k'}^{\rm EX}]
+
i\sum_{i} \sum_{i'} {\bf r}_{i'} [H_{i}^{\rm Z},H_{i'}^{\rm Z}]
\nonumber\\
&+&i\sum_{\langle jk \rangle} \sum_{i'} {\bf r}_{i'} [H_{jk}^{\rm EX},H_{i'}^{\rm Z}]
+
i\sum_{i} \sum_{\langle j'k' \rangle} \frac{{\bf r}_{j'}+{\bf r}_{k'}}{2} [H_{i}^{\rm Z},H_{j'k'}^{\rm EX}].
\nonumber
\end{eqnarray}
In the last equality, the first (second) line shows time reversal odd (even) terms.
Interestingly, time reversal odd terms only arise from the commutator $[H_{jk}^{\rm EX},H_{j'k'}^{\rm EX}]$ because the other commutator is alway zero ($[H_{i}^{\rm Z},H_{i'}^{\rm Z}]=0$).
As an example, explicit calculations for the Kitaev limit lead to the current operator
\begin{equation}
{\bf J}^{\rm E}=\sum_p {\bf j}_p^{\rm E}
\end{equation}
where the local current operator ${\bf j}_p^{\rm E}$ at plaquette $p$ is given by
\begin{eqnarray}
{\bf j}_p^{\rm E}
&=&
\frac{K^2}{2} ({\bf r}_2-{\bf r}_6) S_2^xS_1^zS_6^y 
+ 
\frac{K^2}{2} ({\bf r}_5-{\bf r}_3) S_5^xS_4^zS_3^y 
\nonumber\\
&+&
\frac{K^2}{2} ({\bf r}_6-{\bf r}_4) S_6^zS_5^yS_4^x 
+ 
\frac{K^2}{2} ({\bf r}_3-{\bf r}_1) S_3^zS_2^yS_1^x
\nonumber\\
&+&
\frac{K^2}{2} ({\bf r}_4-{\bf r}_2) S_4^yS_3^xS_2^z
+
\frac{K^2}{2} ({\bf r}_1-{\bf r}_5) S_1^yS_6^xS_5^z 
\nonumber\\
&+&
(\textup{time reversal even terms}).
\end{eqnarray}
Notice the appearance of the components of the chirality operator $\hat{\chi}_p$~\cite{sref_Motome2017}.

\subsection{Validity conditions of the perturbative parton theory and  chirality operator method}

The perturbative parton theory and chirality operator method become exact in the limit of the pure Kitaev model under weak magnetic field. 
Yet, it is important to note that the two methods have their own validity conditions. 
It is exceedingly difficult to prove which one is a better approach in general, and additional careful analysis is necessary if the two methods show discrepancy.
Below, we discuss the validity conditions of each method as well as how to improve the discrepancy. 

Recall that the topological invariant ($\nu$) is determined by the zero temperature limit of thermal hall conductivity over temperature, $\kappa_{xy}/T$.
The perturbative parton theory introduces parameters made of non-Kitaev interactions and flux gap such as $\varGamma'/\Delta_{{\rm flux}}$, and the topological invariant can be written as the expansion, 
\begin{eqnarray}
\nu = {\rm sgn} \left[m_0 + m_1 \left(\frac{\varGamma'}{\Delta_{{\rm flux}}} \right) + m_2 \left( \frac{\varGamma'}{\Delta_{{\rm flux}}} \right)^2 +\cdots \right]. \nonumber
\end{eqnarray}
The coefficients $m_{0,1,\cdots}$ are functions of the coupling constants of a given Hamiltonian ($K, J, \varGamma, \varGamma',\cdots$). 
The perturbative parton theory is valid if the parameters such as $\varGamma'/\Delta_{{\rm flux}}$ are small enough. 
The case \#2 is expected to be similar to the case \#1 as manifested in Fig.~2. For the cases (\#3,4), the perturbation parameters are larger, for example $ \varGamma'/\Delta_{{\rm flux}}=0.77$.
Though the perturbative parton theory is powerful, we also note that the convergence of the expansion is neither guaranteed nor proven, especially with multiple coupling constants.

The chirality operator method, on the other hand, is based on a different type of expansions by exploiting symmetry properties.
Since $\nu$ is in the time reversal odd $A_2$ representation of $D_3$ symmetry, we consider the operator associated with $\kappa_{xy}$, commutator/correlator of energy currents as used in Ref.~\cite{sref_Motome2017}, which can be formally written as 
\begin{equation}
\hat{A}_2 = 
M_3 \sum \hat{\chi}_3 + M_5 \sum \hat{\chi}_5 + M_7 \sum \hat{\chi}_7 + \cdots. \nonumber
\end{equation}
The string operators with an odd number of spin operators connecting two sites on a same sublattice ($\hat{\chi}_{3,5,7}$) are introduced, whose graphical representations are given in Supplementary Fig.~\ref{fig:S4}. Their expectation values ($\langle \hat{\chi}_{3,5,7} \rangle$) determine $\nu$. 
Note that there is no linear term in the expansion as shown in Supplementary Note 7, and the first term $\sum \hat{\chi}_3$ is identical to the chirality operator $\sum_p \hat{\chi}_p$.
The coefficients $M_{3,5,7}$ are functions of the coupling constants of a given Hamiltonian ($K, J, \varGamma, \varGamma',\cdots$) while the string operators are independent. 
The chirality operator method is to truncate the expansion at the leading chirality operator term to determine $\nu$.

The ED calculations with the chirality operator are valid under the conditions, 
\begin{eqnarray}
\Big| M_3  \sum \langle  \hat{\chi}_3 \rangle \Big| \gg  \Big| M_5  \sum  \langle \hat{\chi}_5 \rangle \Big|,  \Big|M_7  \sum  \langle \hat{\chi}_7 \rangle \Big|, \cdots.~~~ \label{validity}
\end{eqnarray}
The conditions hold near the pure Kitaev model as manifested in the perfect matches between the ED calculations and  the perturbative parton theory for the cases (\#1,2).  
The structure of the chirality operator method can be further analyzed by evaluating the expectation values of $\hat{\chi}_{3,5,7}$ for the cases (\#1-4) with a magnetic field along the $[111]$ direction, shown in Supplementary Fig.~\ref{fig:S4}.
We find that the cases (\#3,4) have hundred times larger values of $| \langle \hat{\chi}_n \rangle|$ compared to the cases (\#1,2), which are consistent with the larger perturbation parameters of the perturbative parton theory. 
Moreover, we find that the two ratios,  
\begin{equation}
r_{1}\equiv \frac{|\langle \hat{\chi}_5 \rangle|}{|\langle \hat{\chi}_3 \rangle|}  \sim \frac{1}{10}, \quad \quad  r_{2} \equiv  \frac{|\langle \hat{\chi}_7 \rangle|}{|\langle \hat{\chi}_3 \rangle|} \sim \frac{1}{100} \nonumber
\end{equation}
for the four cases. Since the chirality operator method works well for the cases (\#1,2), the validity conditions [Supplementary Eq.~\eqref{validity}] can be fulfilled if the coefficients $M_{3,5,7}$ of the cases (\#3,4) are not much different  from the ones of the cases (\#1,2).

To check the validity conditions, Supplementary Eq. \eqref{validity}, one needs to calculate the coefficients $M_{3,5,7}$ explicitly. Leaving them for future works, we instead notice that the coefficients $M_{3,5,7}$ are in the trivial representation of the $D_3$ symmetry, being functions of energy eigenvalues. 
We calculate the energy spectrum variances of the lowest hundred energy eigenvalues for the four cases and find that the variances are all less than $5\%$. 
Thus, it is tempting to assume that $M_{3,5,7,\cdots}$ do not vary much, and then, the ED calculations with the chirality operator would work. 

Our proposal with the chirality operator is not only alternative but also complementary to the perturbative parton analysis. 
Namely, the perfect matches between the two methods' results for the cases (\#1,2) become a sanity check  as shown in Fig.~2. 
If the two methods show discrepancy as in the cases (\#3,4), further analysis of the systems are necessary. 
For the perturbative parton analysis, one obvious way is to perform higher order perturbations. 
For the chirality operator method, one can try other numerical methods such as density-matrix-renormalization-group (DMRG) calculations since ED calculations with a larger system size are numerically difficult. 
The topological invariant is believed to be less susceptible to slight  symmetry breaking from a cylinder-like lattice shape, and the chirality operator method is expected to be useful even with DMRG calculations.

\begin{figure}[tb]
\centering
\includegraphics[width=0.7\linewidth]{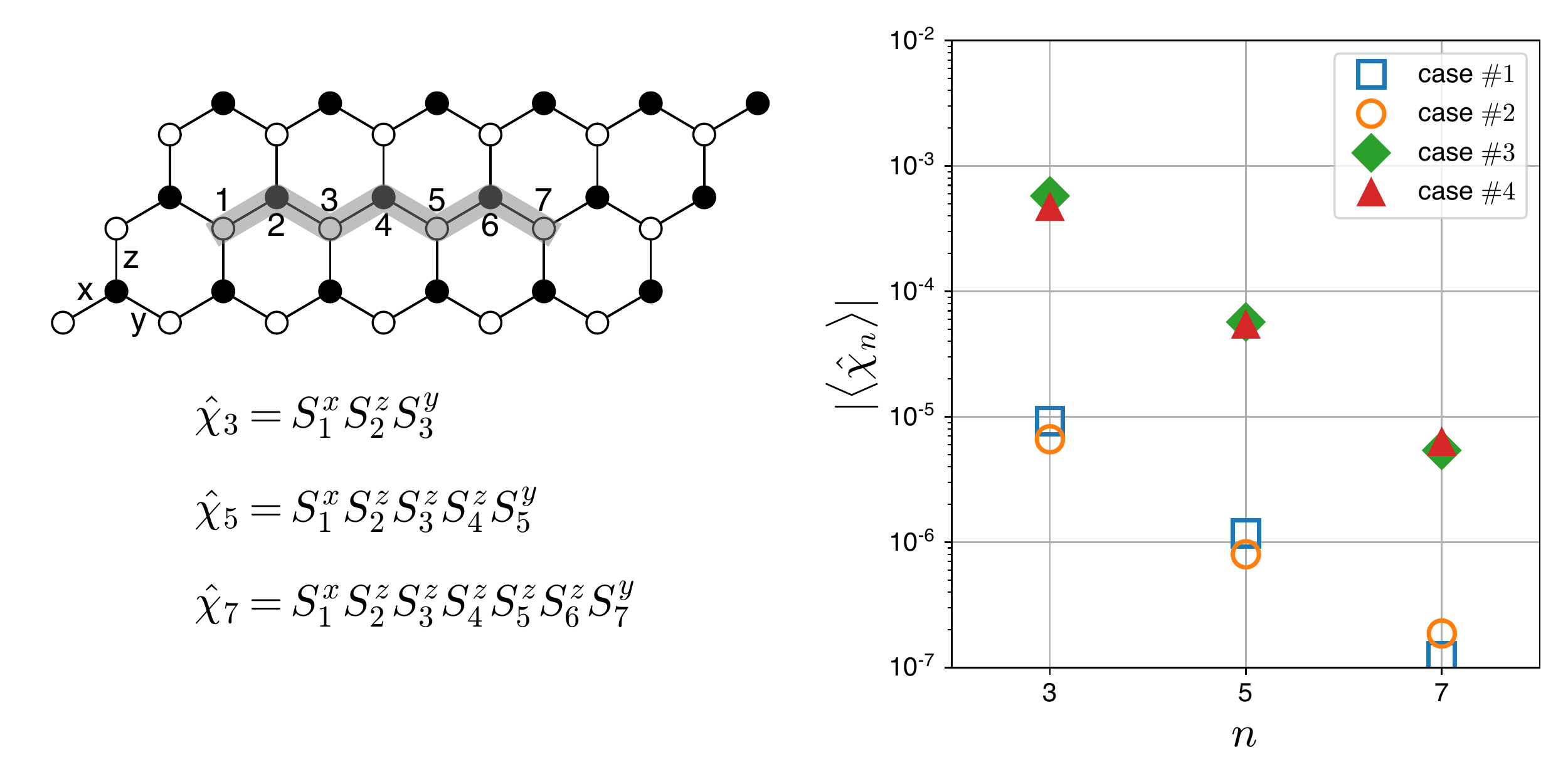}
\renewcommand{\figurename}{\textbf{Supplementary Figure}}
\caption{
{\bf Expectation values of the $A_2$ operators.}
Left: pictorial illustration of the operators $\hat{\chi}_{3,5,7}$.
Right: the expectation values $\langle \hat{\chi}_{3,5,7} \rangle$ for the four cases \#1,2,3,4.
The y-axis is in a log scale, and all the results are obtained by 24-site ED calculations with the field $h=0.01\parallel [111]$.
}
\label{fig:S4}
\end{figure}

\subsection{Spin wave theory}

\begin{figure*}[tb]
\centering
\includegraphics[width=\linewidth]{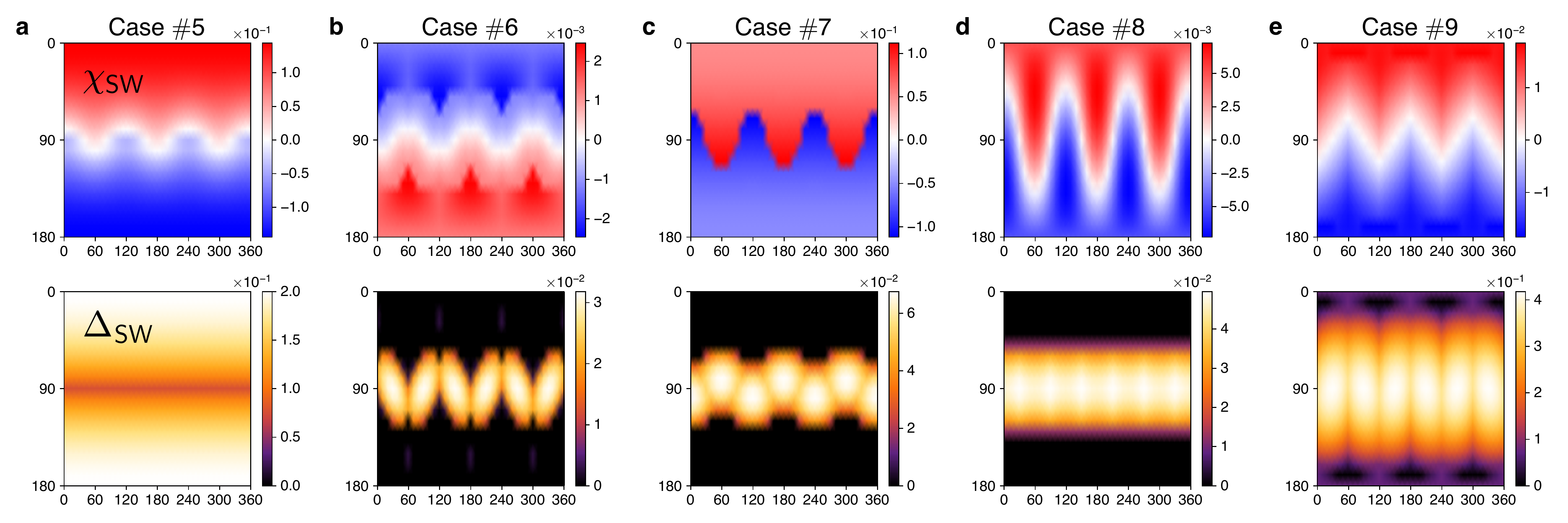} 
\renewcommand{\figurename}{\textbf{Supplementary Figure}}
\caption{{\bf Results of the spin wave theory.}
Upper and bottom: color maps of the chirality $\chi_{\rm SW}({\bf h})$, and the magnon energy gap $\Delta_{\rm SW}({\bf h})$ on the plane of the field angles $(\theta,\phi)$ (horizontal axis: $\phi~[^\circ]$, vertical axis: $\theta~[^\circ]$).
The parameter sets used in the five cases ($\#5\sim9$) are listed in Table 1.
The magnetic field strength is fixed by $h=0.1$ in the cases $\#5\sim8$ and  $h=0.06$ in the case $\#9$.
The black regions in the magnon gap indicate the regions where each magnetic order becomes unstable by magnon condensation.}
\label{fig:S5}
\end{figure*}

\noindent{\bf Magnon gap and chirality.}
Field angle dependence of magnetically ordered phases are investigated using spin wave theories based on the linearized Holstein-Primakoff spin representation:
\begin{equation}
\begin{array}{ccl}
{S}^X &=& \sqrt{\frac{S}{2}} (a+a^{\dagger}),
\\
{S}^Y &=& -i \sqrt{\frac{S}{2}} (a-a^{\dagger}),
\\
{S}^Z &=& S- a^{\dagger} a,
\end{array}
\end{equation}
where the local axis $Z$ is defined by the classical spin configuration $\{{\bf S}_i^{\rm cl}\}$ of the Hamiltonian $H(\theta,\phi)$, and the other two local axes $X,Y$ are perpendicular to the axis $Z$ ($S=1/2$).
The boson operator $a$ describes a local spin-flip, essentially the magnon excitation.
Applying the linearized representation to the Hamiltonian $H(\theta,\phi)$ leads to the quadratic magnon Hamiltonian
\begin{eqnarray}
\mathcal{H}_{\textup{SW}}=E_{\textup{cl}}
&+& \sum_{\bf k}\sum_{m,n=1}^{N_{\rm s}} A_{mn}({\bf k})~a_{m,{\bf k}}^{\dagger}a_{n,{\bf k}}
\nonumber\\
&+& \sum_{\bf k}\sum_{m,n=1}^{N_{\rm s}} B_{mn}({\bf k})~a_{m,-{\bf k}}a_{n,{\bf k}} + \textup{H.c.}
\nonumber\\
&+& \sum_{\bf k} \sum_{m=1}^{N_{\rm s}} ({\bf h} \cdot {\bf S}_m^{\rm cl}) a_{m,{\bf k}}^{\dagger} a_{m,{\bf k}},
\label{eq:H_SW}
\end{eqnarray}
where $E_{\textup{cl}}$ is the energy of the classical spin configuration $\{{\bf S}_i^{\rm cl}\}$, ${N_{\rm s}}$ is the number of sites in the magnetic unit cell, the subscripts $m,n$ denote magnetic sublattices, and ${\bf k}$ represents momentum.
The hopping amplitude $A_{mn}({\bf k})$ and pairing amplitude $B_{mn}({\bf k})$ are determined by the classical spin configuration.
Diagonalizing the magnon Hamiltonian via Bogoliubov transformation, we obtain the magnon energy gap $\Delta_{\rm SW}$ shown in Fig.~4.

For the chirality computation, we simply use the classical spin configuration $\{{\bf S}_i^{\rm cl}\}$ since the chirality should be mainly determined by the local spin moments (for magnetically ordered phases).
We provide the calculated chirality $\chi_{\rm SW}$ together with the magnon gap $\Delta_{\rm SW}$ in Supplementary~Fig.~\ref{fig:S5}.

It is important to note that there is no critical line in the magnon gap.
Anisotropic spin interactions of the Kitaev and Gamma terms and the Zeeman coupling break all continuous spin rotational symmetries, so there is no gapless spin excitation.
Unlike the KQSL, magnetically ordered phases do not show any resemblance/correlation between the excitation energy gap $\Delta_{\rm SW}$ and the chirality $\chi_{\rm SW}$.

\begin{figure}[tb]
\centering
\includegraphics[width=0.5\linewidth]{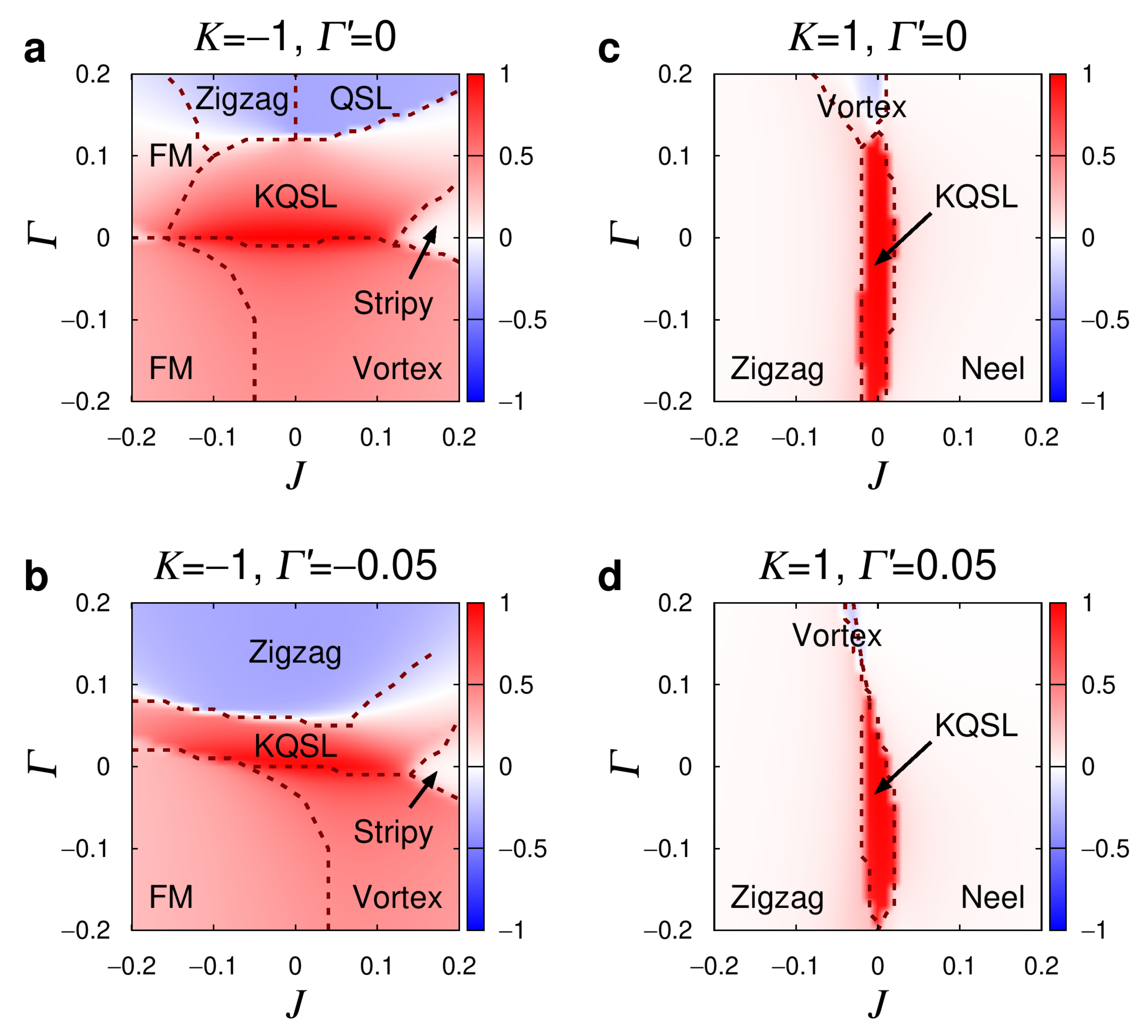} 
\renewcommand{\figurename}{\textbf{Supplementary Figure}}
\caption{{\bf $\boldsymbol{\varGamma'}$ effects on the phases of the $\boldsymbol{K}$-$\boldsymbol{J}$-$\boldsymbol{\varGamma}$-$\boldsymbol{\varGamma'}$ model.}
}
\label{fig:S6}
\end{figure}

\noindent{\bf Zigzag state.}
Based on the ED results, the zigzag antiferromagnetic state occurs in two distinct parameter regimes.
One region is $(K<0,~\varGamma>0,~\varGamma'\leq0)$ in Supplementary~Fig.~\ref{fig:S6}a-b, and the other is $(K>0,~J<0)$ as shown in Fig.~7a and Supplementary~Fig.~\ref{fig:S6}c-d.
We focus on the former case because it is believed to be more relevant to $\alpha$-RuCl$_3$.

At zero field, the classical ground state manifold of $H_{KJ\varGamma\varGamma'}$ is triply degenerate consisting of $x$, $y$, $z$-zigzag states.
These three states are related by $C_3$ rotation, and the name of each state implies the correlation pattern of spin moments.
For instance, in the $z$-zigzag state spin moments are all perpendicular to the $z$-bond axis with being anti-aligned at each $z$-bond.

The degeneracy is lifted when a magnetic field is applied.
The degeneracy lift pattern could be understood most intuitively at in-plane fields ($\theta=90^{\circ}$).
If the magnetic field is bond-parallel, {\it e.g.,} aligned to the $z$-bond axis ($\phi=90^{\circ}$), the $z$-zigzag state becomes most stable compared to the other two states.
In this case, the field is perpendicular to the spin moments of the $z$-zigzag state, which renders the $z$-zigzag state to develop the largest magnetization amongst the three zigzag states.
By a similar mechanism, the $x$-zigzag state is selected near the $x$-bond ($\phi=30^{\circ}$) direction and the $y$-zigzag state is selected near the $y$-bond ($\phi=150^{\circ}$) direction.
From this analysis, we gain the insight that the zigzag phase becomes most stable at the {\it bond-parallel} field directions $\phi=30^{\circ}+n\cdot60^{\circ}~(n=0,1,2,3,4,5)$.
It is corroborated with the largest size of magnon gap $\Delta_{\rm magnon}$ found at the bond-parallel directions (Fig.~5b).

We have checked the same pattern of field angle dependence in the $J$-$K$-$\varGamma$-$J_3$ model proposed for $\alpha$-RuCl$_3$~\cite{sref_Winter2017magnon_breakdown,sref_Winter2018} as well.

\subsection{Topological degeneracy and modular matrix}

\begin{figure*}[tb]
\centering
\includegraphics[width=0.8\linewidth]{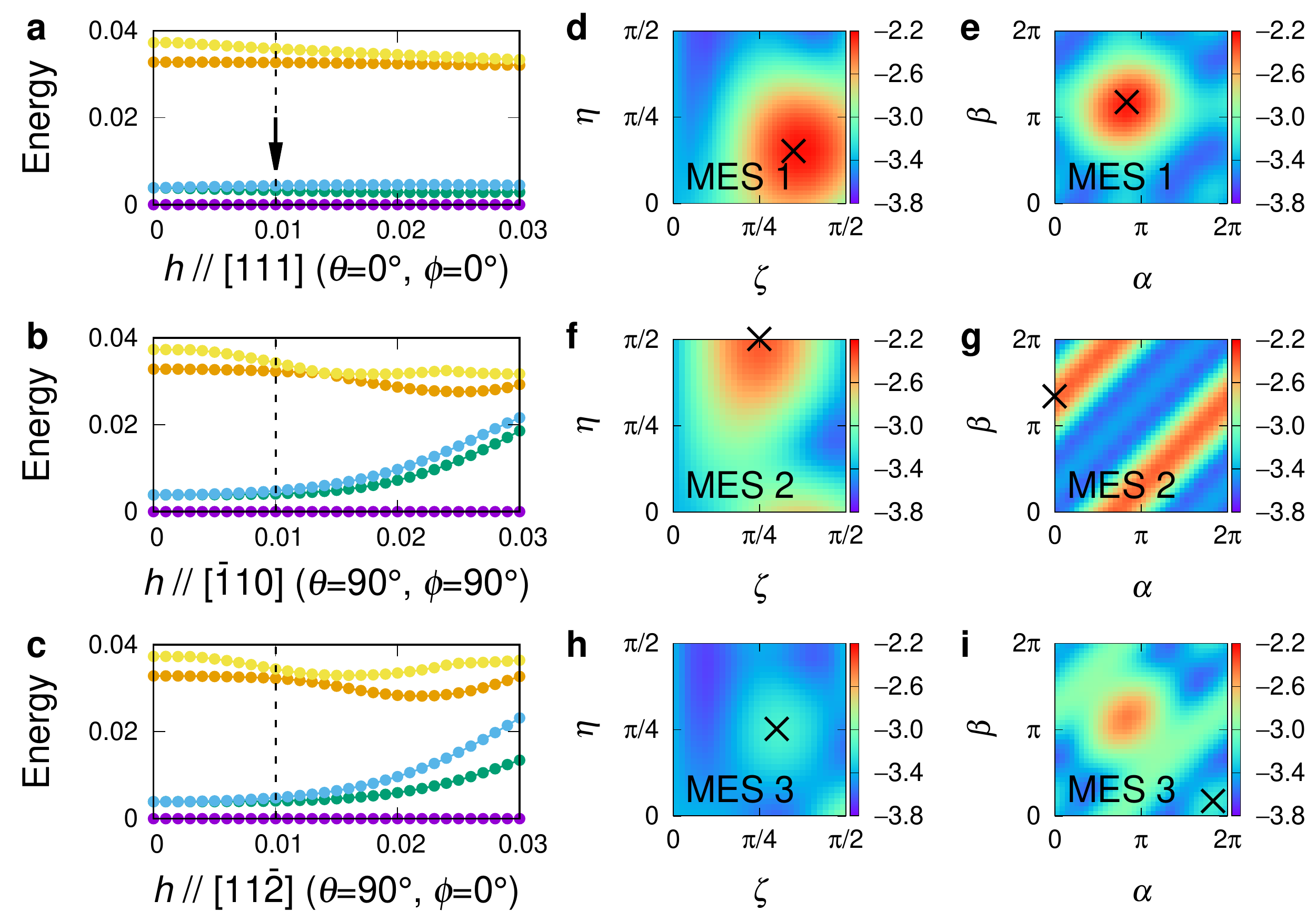} 
\renewcommand{\figurename}{\textbf{Supplementary Figure}}
\caption{
{\bf Topological degeneracy and entanglement entropy.}
{\bf a-c} Evolution of the energy spectrum for three different field directions.
{\bf d-i} Color maps of the entanglement entropy $S_A(\zeta,\eta,\alpha,\beta)$ near the states $\{| \Psi_m^{\rm MES-I} \rangle\}_{m=1}^3$ in Supplementary~Table~\ref{tab:S3}.
In each map, the entropy is plotted with a reversed sign, $-S_A$, which reveals the MES through local maxima.
The cross marks denote the location of $|\Psi_1^{\rm MES-I}\rangle$ in {\bf d-e}, $|\Psi_2^{\rm MES-I}\rangle$ in {\bf f-g}, and $|\Psi_3^{\rm MES-I}\rangle$ in {\bf h-i}.
All the results are obtained for the case~$\#4$ of Table~1, $(K,J,\varGamma,\varGamma')=(-1,0.08,0.01,0.05)$,
and the computations of entanglement entropy and MES are conducted for the magnetic field $h=0.01\parallel[111]$ (marked by an arrow in {\bf a}).
}
\label{fig:S7}
\end{figure*}

\begin{table}[b]
\begin{ruledtabular}
\begin{tabular}{lccccc}
MES & $\frac{\zeta}{\pi/2}$ & $\frac{\eta}{\pi/2}$ & $\frac{\alpha}{2\pi}$
& $\frac{\beta}{2\pi}$ & $S_A$
\\
\hline
$|\Psi_1^{\rm MES-I}\rangle$ & $7/10$ & $3/10$ & $5/12$ & $7/12$ &  2.22
\\
$|\Psi_2^{\rm MES-I}\rangle$ & $5/10$ & $10/10$ & $0/12$ & $8/12$ &  2.36
\\
$|\Psi_3^{\rm MES-I}\rangle$ & $6/10$ & $5/10$ & $11/12$ & $1/12$ &  3.15
\\
\hline
$|\Psi_1^{\rm MES-II}\rangle$ & $7/10$ & $3/10$ & $1/12$ & $11/12$ &  2.22
\\
$|\Psi_2^{\rm MES-II}\rangle$ & $5/10$ & $10/10$ & $0/12$ & $4/12$ &  2.36
\\
$|\Psi_3^{\rm MES-II}\rangle$ & $6/10$ & $5/10$ & $7/12$ & $5/12$ &  3.15
\end{tabular}
\end{ruledtabular}
\renewcommand{\tablename}{\textbf{Supplementary Table}}
\caption{
{\bf Minimally entangled states.} 
The MES are obtained by searching on a $10\times10\times12\times12$ uniform grid of $(\zeta,\eta,\alpha,\beta)$ for the case $\#4$ of Table 1, $(K,J,\varGamma,\varGamma')=(-1,0.08,0.01,0.05)~~\&~~h=0.01\parallel[111]$.
}
\label{tab:S3}
\end{table}

The non-abelian KQSL state has threefold ground state degeneracy on a torus geometry due to the Ising anyon topological order~\cite{sref_Kitaev,sref_topo_deg_Kells2009}. In Supplementary~Fig.~\ref{fig:S7}, we demonstrate the degeneracy for a few selected field directions.
It is clearly shown that the lowest three states (degeneracy slightly lifted due to the finite size effect) are well separated from the other excited states for small $h$. 
Furthermore, it is verified that those quasi-degenerate states share qualitatively same bulk properties such as the expectation value of flux operator $\langle \hat{W}_p \rangle$, spin structure factor $S({\bf q})$, and also chirality $\chi({\bf h})$.

By using the topological degenerate states, we may extract the modular $\mathcal{S}$ matrix containing the information of quasiparticles' statistics and fusion rules.
It is achieved by finding the so called minimally entangled states (MES) in the subspace of the quasi-degenerate states~\cite{sref_MES_Zhang2012,sref_Sheng2014}:
\begin{equation}
| \Psi \rangle = \sum_{n=1}^3 z_n | \Psi_n \rangle,
\end{equation}
where $| \Psi_{1,2,3} \rangle$ are the three quasi-degenerate states, and the complex coefficients $z_{1,2,3}$ are parametrized by
\begin{eqnarray}
z_1&=&\sin \zeta \cos \eta,
\\
z_2&=&\sin \zeta \sin \eta e^{i\alpha},
\\
z_3&=&\cos \zeta e^{i\beta},
\end{eqnarray}
with the four angles, $\zeta,\eta \in [0,\pi/2]$ and $\alpha,\beta \in [0, 2\pi]$.

To construct the minimally entangled states, we consider two noncontractable bipartitions on the torus geometry (cut I and cut II)~~\cite{sref_MES_Zhang2012,sref_Sheng2014}.
Then for each bipartition we compute the entanglement entropy
\begin{equation}
S_A  = - \log {\rm Tr} \rho_A^2
\end{equation}
where $\rho_A$ is the reduced density matrix $\rho_A \equiv {\rm Tr}_B | \Psi \rangle \langle \Psi | $ traced over the partition $B$ of the torus.
The minimally entangled states $\{|\Psi_n^{\rm MES}\rangle\}_{n=1}^3$ correspond to the local minima of $S_A$ in the parameter space of $(\zeta,\eta,\alpha,\beta)$.
We list the resulting MES in Supplementary~Table~\ref{tab:S3}, and 
illustrate the entanglement entropy $S_A(\zeta,\eta,\alpha,\beta)$ near the MES points in Supplementary~Fig.~\ref{fig:S7}d-i.
After an appropriate U(1) transformation in each of the MES,
the inner products between the two sets of the MES
\begin{equation}
\mathcal{S}_{mn}=\langle \Psi_m^{\rm MES-I} | \Psi_n^{\rm MES-II} \rangle
\end{equation}
yield the modular $\mathcal{S}$ matrix of the Ising topological field theory in Eq.~(7).

\begin{figure*}[tb]
\centering
\includegraphics[width=\linewidth]{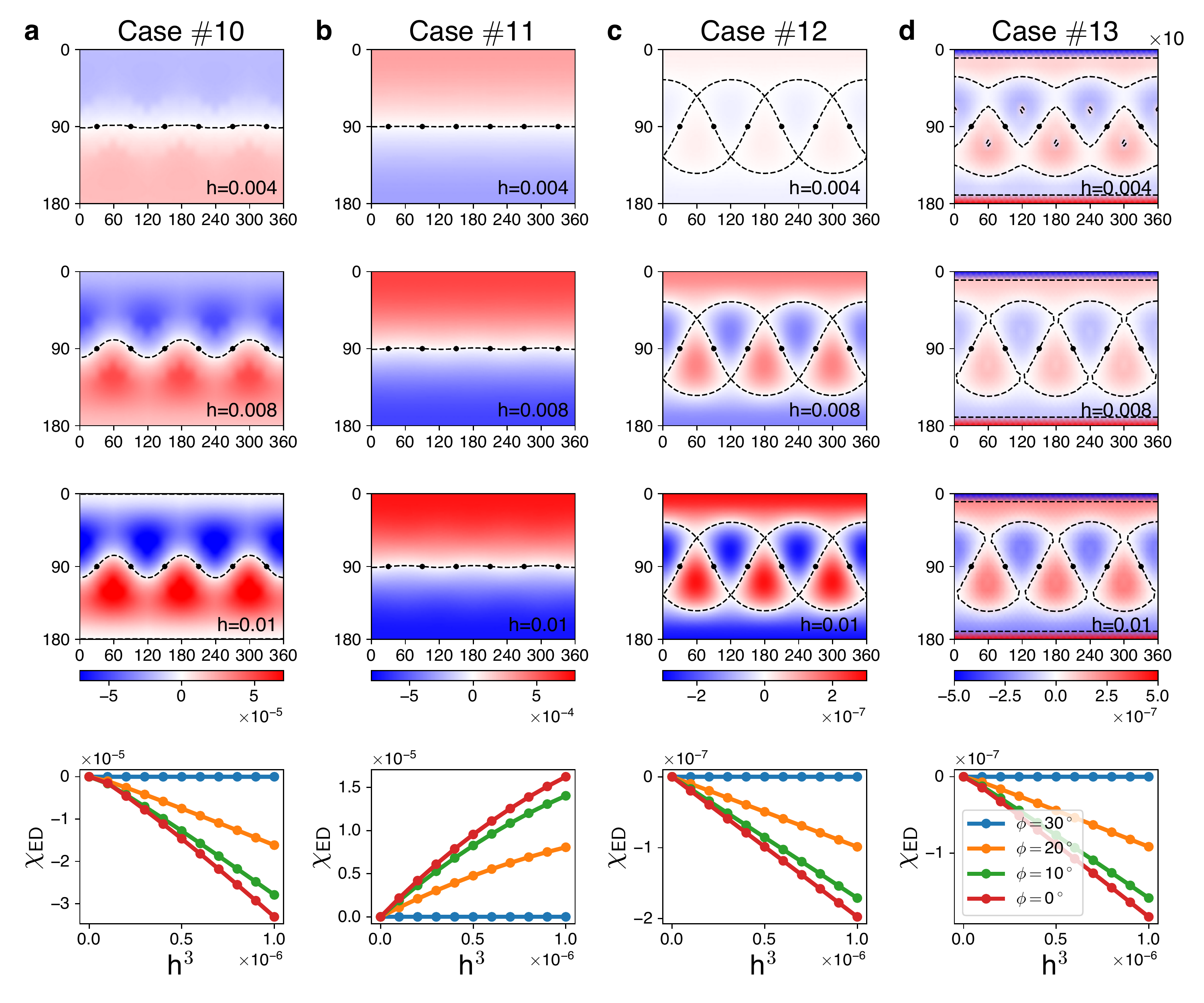} 
\renewcommand{\figurename}{\textbf{Supplementary Figure}}
\caption{{\bf Additional results of the chirality for the non-abelian KQSL.}
Top three: color maps of the chirality $\chi_{\rm ED}({\bf h})$ on the plane of the field angles $(\theta,\phi)$ for the magnetic field strength $h=0.004,~0.008,~0.01$ (horizontal axis: $\phi~[^\circ]$, vertical axis: $\theta~[^\circ]$).
The dashed lines highlight the zero lines $\chi_{\rm ED}({\bf h})=0$, and the black dots mark the bond directions.
Bottom: $\chi_{\rm ED}({\bf h})$ as a function of $h^3$ for the in-plane fields $(\theta=90^\circ,\phi=0^\circ,10^\circ,20^\circ,30^\circ)$, illustrating the universality of the $h^3$ behavior in the KQSL.
The parameter sets used in the four cases ($\#10\sim13$) are listed in Supplementary~Table~\ref{tab:S4}.
}
\label{fig:S8}
\end{figure*}

\subsection{Additional results of the chirality $\chi_{\rm ED}$}

Supplementary~Fig.~\ref{fig:S8} presents additional results of the chirality $\chi_{\rm ED}$,
obtained for the parameter sets in Supplementary~Table~\ref{tab:S4}.
The first three cases $\#10\sim12$ are well understood by the competition between the $h$-linear term $(h_x+h_y+h_z)$ and the $h$-cubic term $(h_xh_yh_z)$.
By contrast, the case $\#13$ shows distinguished field evolution behaviors from the other cases.
For instance, high intensity peaks of $\chi_{\rm ED}$ are observed near the two poles $\theta=0^\circ,~180^\circ$, enclosed by additional critical lines.
These features are attributed to effects of other $h$-cubic terms such as $F'_3({\bf h})$ and $F''_3({\bf h})$ in Supplementary~Table~\ref{tab:A2-expansion}.

The cubic dependence for in-plane fields is confirmed in most of the cases.
We find a slight deviation from the cubic dependence in the case $\#10$ near the zero field.
To confirm the cubic dependence in ED calculations, it is important to make sure that topological degeneracy is lifted by a small but finite energy gap, eliminating the degeneracy in the ground state (as in Supplementary~Fig.~\ref{fig:S7}a-c).

\begin{table}[tb]
\begin{ruledtabular}
\begin{tabular}{lcccccc}
Case  &  $K$  &  $J$  &  $\varGamma$  &  $\varGamma'$  &  Phase & Supplementary
\\
 & & & & & & figure
\\
\hline
\#10  & -1  & 0 & 0 & 0.0005 &  KQSL & \ref{fig:S8}a
\\
\#11  & -1 & 0 & 0.01 & -0.01 &  KQSL & \ref{fig:S8}b 
\\
\#12  & 1  & 0 & 0 & 0 &  KQSL & \ref{fig:S8}c
\\
\#13  & 1  & 0 & 0.05 & 0 &  KQSL & \ref{fig:S8}d
\end{tabular}
\end{ruledtabular}
\renewcommand{\tablename}{\textbf{Supplementary Table}}
\caption{{\bf Additional parameter sets for exact diagonalization.}
}
\label{tab:S4}
\end{table}

\subsection{Identification of phase boundaries by the chirality}

Distinct phase boundaries under magnetic field can be identified clearly by the chirality operator.
We demonstrate this by conducting additional ED calculations as shown in Supplementary~Fig.~\ref{fig:S9}. 
We find that the phase boundaries revealed by the plaquette operator ($\hat{W}_p$) are equally well captured by the chirality operator ($\hat{\chi}_p$). 
In some cases, the chirality operator works better than the plaquette operator as shown in Supplementary~Fig.~\ref{fig:S9}b,e (marked by arrows).
This shows that the chirality is also useful for the identification of distinct phase boundaries of the Kitaev system.

\begin{figure*}[tb]
\centering
\includegraphics[width=\linewidth]{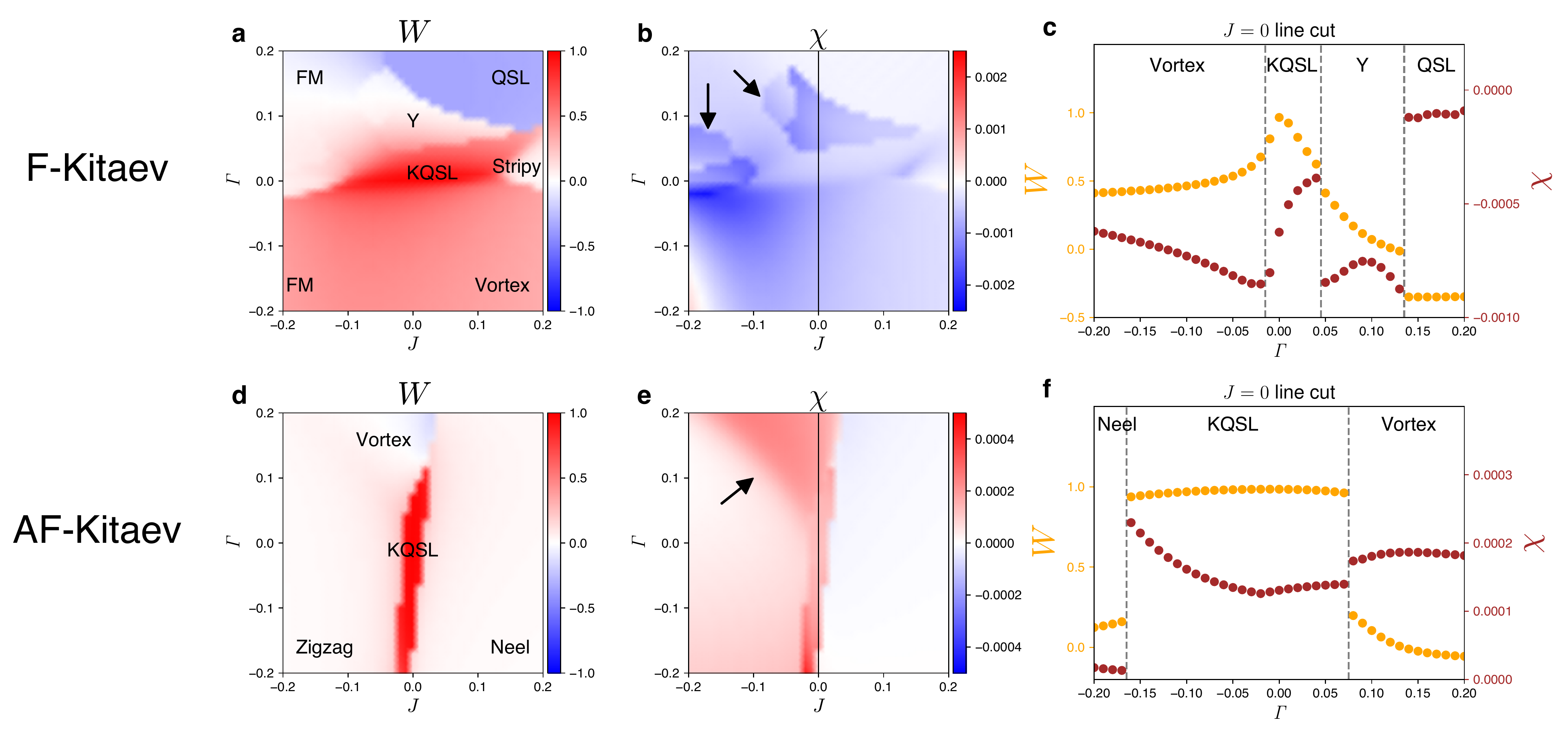}
\renewcommand{\figurename}{\textbf{Supplementary Figure}}
\caption{
{\bf Distinct phase boundaries in the chirality.}
{\bf a-c} ED results for $K=-1$, $\varGamma'=0.05$, $h=0.01\parallel [111]$ with the flux $W~(=\langle \hat{W}_p \rangle)$, the chirality $\chi~(=\langle \hat{\chi}_p \rangle)$, and a comparison of $W$ and $\chi$ along the $J=0$ line cut.
{\bf d-f} ED results for $K=1$, $\varGamma'=-0.05$, $h=0.01\parallel [111]$ with the flux $W$, the chirality $\chi$, and a comparison of $W$ and $\chi$ along the $J=0$ line cut.
}
\label{fig:S9}
\end{figure*}



\end{document}